\documentclass[acmsmall,screen,nonacm]{acmart}

\usepackage{graphicx}
\usepackage{threeparttable}
\usepackage{array}
\usepackage{xcolor}
\usepackage{tikz}
\usetikzlibrary{positioning,fit,arrows.meta,backgrounds,calc}
\usepackage{pgfplots}
\pgfplotsset{compat=1.18}
\usepgfplotslibrary{groupplots}
\definecolor{mOpusFive}{HTML}{2A78D6}
\definecolor{mFable}{HTML}{EB6834}
\definecolor{mSol}{HTML}{1BAF7A}
\definecolor{mAstra}{HTML}{EDA100}
\pgfplotsset{libgit2 campaign axis/.style={xtick={0,2,4,7,10,13,17,24,27}, xticklabels={void/string,util,oid,core,repository,object,config,index,modules}, xmin=-0.6, xmax=33.6, xticklabel style={{rotate=50, anchor=north east, inner sep=1pt}}}}

\pgfplotsset{ffmpeg campaign axis/.style={xtick={0,2,5,10,13,15,20,23}, xticklabels={void/string,core,formats,options,crypto,buffer-frame,hwcontext,modules}, xmin=-0.6, xmax=25.6, xticklabel style={{rotate=50, anchor=north east, inner sep=1pt}}}}

\pgfplotsset{
  batchplot/.style={
    width=0.52\linewidth, height=3.4cm,
    tick label style={font=\scriptsize}, label style={font=\footnotesize},
    title style={font=\footnotesize, yshift=-1ex},
    axis lines*=left, xmajorgrids, ymajorgrids, grid style={black!8},
    scatter, only marks, mark size=1.6pt,
    point meta=explicit symbolic,
    unbounded coords=discard,
    table/col sep=comma,
  },
  only rows/.style 2 args={
    x filter/.append code={%
      \edef\batchrowval{\thisrow{#1}}\edef\batchwant{#2}%
      \ifx\batchrowval\batchwant\else\fi}},
  filled models/.style={scatter/classes={
    opus55={mark=*, draw=mOpusFive, fill=mOpusFive},
    fable51={mark=square*, draw=mFable, fill=mFable},
    sol61={mark=triangle*, draw=mSol, fill=mSol},
    astra6={mark=diamond*, draw=mAstra, fill=mAstra}}},
  hollow models/.style={scatter/classes={
    opus55={mark=*, draw=mOpusFive, fill=white, line width=0.7pt},
    fable51={mark=square*, draw=mFable, fill=white, line width=0.7pt},
    sol61={mark=triangle*, draw=mSol, fill=white, line width=0.7pt},
    astra6={mark=diamond*, draw=mAstra, fill=white, line width=0.7pt}}},
  batch legend/.style={legend columns=-1,
    legend style={font=\scriptsize, text=black, draw=none, column sep=0.6ex}},
}
\newcommand{\batchlegend}{%
  \addlegendimage{only marks, mark=*, draw=mOpusFive, fill=mOpusFive}\addlegendentry{Opus 5.5}
  \addlegendimage{only marks, mark=square*, draw=mFable, fill=mFable}\addlegendentry{Fable 5.1}
  \addlegendimage{only marks, mark=triangle*, draw=mSol, fill=mSol}\addlegendentry{GPT-6.1 Sol}
  \addlegendimage{only marks, mark=diamond*, draw=mAstra, fill=mAstra}\addlegendentry{GPT-6 Astra}}
\newcommand{\batchcost}[3]{%
  \addplot[filled models, only rows={kind}{#3}] table[x=t, y=usd, meta=model] {#1};
  \addplot[hollow models, only rows={kind}{#3}] table[x=t, y=usd, meta=model] {#2};}
\newcommand{\batchsize}[1]{%
  \addplot[filled models, only rows={kind}{#1}] table[x=xs, y=usd, meta=model] {data/libgit2-wrap.csv};
  \addplot[filled models, only rows={kind}{#1}] table[x=xs, y=usd, meta=model] {data/ffmpeg-wrap.csv};
  \addplot[hollow models, only rows={kind}{#1}] table[x=xs, y=usd, meta=model] {data/libgit2-review.csv};
  \addplot[hollow models, only rows={kind}{#1}] table[x=xs, y=usd, meta=model] {data/ffmpeg-review.csv};}

\usepackage[nonumberlist]{glossaries}
\glsdisablehyper
\setacronymstyle{long-short}
\newacronym{asan}{ASan}{AddressSanitizer}
\newacronym{ci}{CI}{continuous integration}
\newacronym{ffi}{FFI}{foreign function interface}
\newacronym{hir}{HIR}{high-level intermediate representation}
\newacronym{llm}{LLM}{large language model}
\newacronym{mir}{MIR}{mid-level intermediate representation}
\newacronym{msan}{MSan}{MemorySanitizer}
\newacronym{scc}{SCC}{strongly connected component}
\newacronym{tsan}{TSan}{ThreadSanitizer}
\newacronym{ub}{UB}{undefined behavior}
\newacronym{ubsan}{UBSan}{UndefinedBehaviorSanitizer}

\newcommand{\lgApiTypes}{165}  %
\newcommand{\lgApiSyms}{981}  %
\newcommand{\lgCostUSD}{2{,}308}  %
\newcommand{\lgAgentsCostUSD}{1{,}823}  %
\newcommand{\lgOrchCostUSD}{485}  %
\newcommand{\lgTokensB}{5.1}  %
\newcommand{\lgApiCov}{41.9}  %
\newcommand{\lgApiItems}{1{,}146}  %
\newcommand{\lgWrappedItems}{480}  %
\newcommand{\lgWrappedTypes}{69}
\newcommand{\lgWrappedSyms}{411}
\newcommand{\lgApiCovTypes}{41.8}  %
\newcommand{\lgApiCovSyms}{41.9}  %
\newcommand{\lgTests}{2{,}064}  %
\newcommand{\lgTestsUbSafe}{688}
\newcommand{\lgTestsUbUnsafe}{100}
\newcommand{\lgTestsEquiv}{489}
\newcommand{\lgTestsUnit}{787}
\newcommand{\lgCovRust}{98.92}      %
\newcommand{\lgCovC}{45.95}
\newcommand{\lgUbCovRust}{79.28}    %
\newcommand{\lgUbCovC}{28.10}
\newcommand{\lgUbUnsafeCovRust}{12.81}
\newcommand{\lgUbUnsafeCovC}{22.12}
\newcommand{\lgEqCovRust}{75.93}
\newcommand{\lgEqCovC}{43.47}
\newcommand{\lgUnitCovRust}{87.34}
\newcommand{\lgUnitCovC}{32.03}
\newcommand{\lgCovCCmp}{46.78}  %
\newcommand{\lgRvAddTests}{153}       %
\newcommand{\lgRvAddTestsPct}{7.4}    %
\newcommand{\lgRvAddCovRust}{1.55}    %
\newcommand{\lgRvAddCovC}{0.52}
\newcommand{\lgRvAddUnitCovRust}{3.82} %

\newcommand{\lgPubFns}{2{,}663}
\newcommand{\lgUnsafeFnPub}{102}
\newcommand{\lgUnsafeFnPct}{3.8}  %
\newcommand{\lgUnsafeFnTrait}{55}  %
\newcommand{\lgUnsafeFnSeam}{5}  %
\newcommand{\lgUnsafeFnOther}{42}
\newcommand{\lgUnsafeFnBackend}{3}  %
\newcommand{\lgUnsafeFnOdb}{2}  %
\newcommand{\lgUnsafeFnLift}{2}  %
\newcommand{\lgUnsafeFnOtherPct}{1.6}  %
\newcommand{\lgUnsafeFnGlobal}{35}  %
\newcommand{\lgRawPtrFns}{2}  %
\newcommand{\lgRawPtrFnsPct}{0.1}  %
\newcommand{\lgRawPtrFnsSafeTake}{0}     %
\newcommand{\lgPubWrapFns}{455}
\newcommand{\lgUnsafeWrapFns}{39}
\newcommand{\lgUnsafeWrapFnsPct}{8.6}  %
\newcommand{\lgRawPtrWrapFns}{2}
\newcommand{\lgRawPtrWrapFnsPct}{0.4}  %

\newcommand{\lgTyFfiboxPct}{66.8}  %
\newcommand{\lgTyDrop}{5}
\newcommand{\lgTyDropPct}{1.3}  %
\newcommand{\lgTyRefForming}{0}
\newcommand{\lgTyRawField}{3}  %
\newcommand{\lgTyRawFieldPct}{0.8}  %

\newcommand{\ffApiTypes}{210}  %
\newcommand{\ffApiSyms}{1{,}128}  %
\newcommand{\ffCostUSD}{2{,}120}  %
\newcommand{\ffAgentsCostUSD}{1{,}678}  %
\newcommand{\ffOrchCostUSD}{442}  %
\newcommand{\lgAsstOrchUSD}{249}
\newcommand{\lgAsstPlanUSD}{149}
\newcommand{\lgAsstDevUSD}{86}  %
\newcommand{\ffAsstOrchUSD}{272}
\newcommand{\ffAsstPlanUSD}{31}
\newcommand{\ffAsstDevUSD}{139}  %
\newcommand{\ffTokensB}{4.5}  %
\newcommand{\ffApiCov}{42.8}  %
\newcommand{\ffApiItems}{1{,}338}  %
\newcommand{\ffWrappedItems}{573}  %
\newcommand{\ffWrappedTypes}{57}
\newcommand{\ffWrappedSyms}{516}
\newcommand{\ffApiCovTypes}{27.1}  %
\newcommand{\ffApiCovSyms}{45.7}  %
\newcommand{\ffTests}{2{,}694}  %
\newcommand{\ffTestsUbSafe}{821}
\newcommand{\ffTestsUbUnsafe}{182}
\newcommand{\ffTestsEquiv}{723}
\newcommand{\ffTestsUnit}{968}
\newcommand{\ffCovRust}{99.07}
\newcommand{\ffCovC}{67.20}
\newcommand{\ffUbCovRust}{75.04}    %
\newcommand{\ffUbCovC}{53.10}
\newcommand{\ffUbUnsafeCovRust}{28.16}
\newcommand{\ffUbUnsafeCovC}{26.64}
\newcommand{\ffEqCovRust}{80.38}
\newcommand{\ffEqCovC}{66.67}
\newcommand{\ffUnitCovRust}{84.45}
\newcommand{\ffUnitCovC}{41.20}
\newcommand{\ffRvAddTests}{165}       %
\newcommand{\ffRvAddTestsPct}{6.1}    %
\newcommand{\ffRvAddCovRust}{1.09}
\newcommand{\ffRvAddCovC}{0.42}       %
\newcommand{\ffPubFns}{2{,}637}
\newcommand{\ffUnsafeFnPub}{63}
\newcommand{\ffUnsafeFnPct}{2.4}  %
\newcommand{\ffUnsafeFnTrait}{45}        %
\newcommand{\ffUnsafeFnSeam}{7}  %
\newcommand{\ffUnsafeFnOther}{11}  %
\newcommand{\ffUnsafeFnStd}{2}  %
\newcommand{\ffUnsafeFnCRec}{7}  %
\newcommand{\ffUnsafeFnLift}{2}  %
\newcommand{\ffUnsafeFnOtherPct}{0.4}  %
\newcommand{\ffRawPtrFns}{16}  %
\newcommand{\ffRawPtrFnsPct}{0.6}  %
\newcommand{\ffRawPtrFnsOpaque}{14}  %
\newcommand{\ffRawPtrFnsSafeTake}{7}  %
\newcommand{\ffPubWrapFns}{560}
\newcommand{\ffUnsafeWrapFns}{5}
\newcommand{\ffUnsafeWrapFnsPct}{0.9}  %
\newcommand{\ffRawPtrWrapFns}{2}
\newcommand{\ffRawPtrWrapFnsPct}{0.4}  %

\newcommand{\ffTyFfiboxPct}{66.2}  %
\newcommand{\ffTyDrop}{3}
\newcommand{\ffTyDropPct}{0.9}  %
\newcommand{\ffTyRefForming}{0}
\newcommand{\ffTyRawField}{14}  %
\newcommand{\ffTyRawFieldPct}{4.2}  %

\newcommand{\lgOwnFp}{3}
\newcommand{\ffOwnFp}{11}
\newcommand{\lgOwnFpPct}{0.2}    %
\newcommand{\ffOwnFpPct}{1.2}    %
\newcommand{\lgOwnFnPos}{42}
\newcommand{\ffOwnFnPos}{35}
\newcommand{\lgOwnFnPosPct}{3.3}  %
\newcommand{\ffOwnFnPosPct}{3.9}  %
\newcommand{\lgOwnFnLife}{24}     %
\newcommand{\ffOwnFnLife}{14}     %
\newcommand{\lgOwnFnLifePct}{10.3} %
\newcommand{\ffOwnFnLifePct}{10.1} %
\newcommand{\gitAuditVersion}{0.21.0}
\newcommand{\gitAuditWallHours}{1.4}
\newcommand{\gitAuditCostUSD}{44}
\newcommand{\gitAuditFiles}{19}
\newcommand{\gitAuditFilesAll}{64}
\newcommand{\gitAuditUb}{25}
\newcommand{\gitAuditUbAsan}{15}
\newcommand{\gitAuditUbTsan}{4}
\newcommand{\gitAuditUbBsan}{4}
\newcommand{\gitAuditUbMiri}{2}
\newcommand{\lgReportsAll}{178}
\newcommand{\lgReportsWavefront}{33}
\newcommand{\lgReportsUb}{8}          %
\newcommand{\lgReportsUbSafe}{6}
\newcommand{\lgReportsUbUnsafe}{2}
\newcommand{\lgReportsEquiv}{6}
\newcommand{\lgReportsUnsafe}{7}       %
\newcommand{\lgReportsObligations}{11}   %
\newcommand{\lgReportsOther}{146}      %
\newcommand{\lgUbOob}{5}
\newcommand{\ffUbOob}{2}
\newcommand{\lgUbUaf}{1}
\newcommand{\ffUbUaf}{3}
\newcommand{\ffUbDf}{1}
\newcommand{\lgUbNull}{1}
\newcommand{\ffUbArith}{6}
\newcommand{\ffReportsAll}{138}
\newcommand{\ffReportsWavefront}{34}
\newcommand{\ffReportsUb}{13}         %
\newcommand{\ffReportsUbSafe}{4}
\newcommand{\ffReportsUbUnsafe}{9}
\newcommand{\ffReportsEquiv}{4}
\newcommand{\ffReportsUnsafe}{8}       %
\newcommand{\ffReportsObligations}{8}    %
\newcommand{\ffReportsOther}{105}      %
\newcommand{\lgRvSurface}{3}           %
\newcommand{\ffRvSurface}{6}           %
\newcommand{\lgRvInternal}{4}
\newcommand{\ffRvInternal}{2}
\newcommand{\lgRvUnsafeFnRemoved}{3}
\newcommand{\ffRvUnsafeFnRemoved}{5}
\newcommand{\ffRvRawPtrRemoved}{3}
\newcommand{\lgLocInserted}{165{,}289} %
\newcommand{\ffLocInserted}{101{,}056} %
\newcommand{\lgCodeLines}{17{,}571}   %
\newcommand{\ffCodeLines}{18{,}681}   %
\newcommand{\lgCodeLinesPct}{10.6}     %
\newcommand{\ffCodeLinesPct}{18.5}   %
\newcommand{\lgUnsafeLocPct}{16.6}  %
\newcommand{\ffUnsafeLocPct}{16.6}  %
\newcommand{\lgUnwrappedFns}{487}
\newcommand{\ffUnwrappedFns}{607}
\newcommand{\lgSeamPos}{433}          %
\newcommand{\ffSeamPos}{91}           %
\newcommand{\lgSeamFld}{46}           %
\newcommand{\ffSeamFld}{55}           %
\newcommand{\lgSeamPosRepo}{126}      %
\newcommand{\ffSeamPosFrame}{42}      %
\newcommand{\lgValPos}{15}            %
\newcommand{\ffValPos}{49}            %

\newcommand{\lgBsanSkip}{16}
\newcommand{\lgBsanSkipSlow}{11}
\newcommand{\lgBsanSkipHang}{1}
\newcommand{\lgBsanSkipFp}{4}

\newcommand{\lgMiriRun}{598}
\newcommand{\lgMiriRunPct}{29.0}      %
\newcommand{\ffMiriRun}{700}
\newcommand{\ffMiriRunPct}{26.0}      %

\newcommand{\lgOwnPos}{1{,}262}
\newcommand{\lgOwnOwned}{145}
\newcommand{\lgOwnBorrowed}{1{,}130}
\newcommand{\lgOwnBorrowArgPct}{73}
\newcommand{\lgOwnMutablePct}{43}      %
\newcommand{\lgOwnNullablePct}{34}     %
\newcommand{\lgOwnString}{270}       %
\newcommand{\lgOwnArray}{47}         %

\newcommand{\lgOwnCtors}{150}
\newcommand{\lgOwnDroppers}{54}
\newcommand{\lgOwnCloners}{19}         %
\newcommand{\lgOwnRefcount}{7}
\newcommand{\lgOwnLocked}{2}

\newcommand{\ffOwnPos}{897}
\newcommand{\ffOwnOwned}{118}
\newcommand{\ffOwnBorrowed}{791}
\newcommand{\ffOwnBorrowArgPct}{82}
\newcommand{\ffOwnMutablePct}{46}      %
\newcommand{\ffOwnNullablePct}{42}     %
\newcommand{\ffOwnString}{209}       %
\newcommand{\ffOwnArray}{160}        %

\newcommand{\ffOwnCtors}{100}
\newcommand{\ffOwnDroppers}{20}
\newcommand{\ffOwnCloners}{14}         %
\newcommand{\ffOwnRefcount}{2}
\newcommand{\ffOwnLocked}{3}

\newcommand{\gitApiCov}{65.1}  %
\newcommand{\gitTests}{238}         %
\newcommand{\gitCovRust}{60.59}     %
\newcommand{\gitCovC}{45.12}        %
\newcommand{\fttApiCov}{18.7}  %
\newcommand{\nextApiCov}{17.6}  %
\newcommand{\fttTests}{47}          %
\newcommand{\nextTests}{48}         %
\newcommand{\fttCovRust}{11.34}     %
\newcommand{\nextCovRust}{18.73}     %
\newcommand{\fttCovC}{6.77}         %
\newcommand{\nextCovC}{10.08}       %
\newcommand{\gitPubFns}{2{,}719}

\newcommand{\gitUnsafeFnPct}{4.1}
\newcommand{\gitUnsafeFnOther}{23}      %
\newcommand{\gitUnsafeFnOtherPct}{0.8}
\newcommand{\gitRawPtrFns}{0}           %
\newcommand{\gitRawPtrFnsPct}{0.0}
\newcommand{\gitRawPtrFnsSafeTake}{0}

\newcommand{\gitTyDrop}{52}
\newcommand{\gitTyDropPct}{28.3}
\newcommand{\gitTyRefForming}{3}        %
\newcommand{\gitTyRefFormingPct}{1.6}
\newcommand{\gitTyRawField}{70}
\newcommand{\gitTyRawFieldPct}{38.0}  %

\newcommand{\fttPubFns}{2{,}222}

\newcommand{\fttUnsafeFnPct}{3.7}
\newcommand{\fttUnsafeFnOther}{9}       %
\newcommand{\fttUnsafeFnOtherPct}{0.4}
\newcommand{\fttRawPtrFns}{7}           %
\newcommand{\fttRawPtrFnsPct}{0.3}
\newcommand{\fttRawPtrFnsSafeTake}{6}

\newcommand{\fttTyDrop}{15}
\newcommand{\fttTyDropPct}{8.5}
\newcommand{\fttTyRefForming}{20}       %
\newcommand{\fttTyRefFormingPct}{11.3}
\newcommand{\fttTyRawField}{48}
\newcommand{\fttTyRawFieldPct}{27.1}  %

\newcommand{\nextPubFns}{1{,}955}

\newcommand{\nextUnsafeFnPct}{5.7}
\newcommand{\nextUnsafeFnOther}{11}     %
\newcommand{\nextUnsafeFnOtherPct}{0.6}
\newcommand{\nextRawPtrFns}{13}         %
\newcommand{\nextRawPtrFnsPct}{0.7}
\newcommand{\nextRawPtrFnsSafeTake}{11} %

\newcommand{\nextTyDrop}{17}
\newcommand{\nextTyDropPct}{10.6}
\newcommand{\nextTyRefForming}{26}      %
\newcommand{\nextTyRefFormingPct}{16.1}
\newcommand{\nextTyRawField}{49}
\newcommand{\nextTyRawFieldPct}{30.4}  %

\newcommand{\lgSysFns}{501}

\newcommand{\lgSysFnsWrapped}{435}
\newcommand{\lgSysFnsWrappedPct}{86.8}  %
\newcommand{\ffSysFns}{528}            %

\newcommand{\ffSysFnsWrapped}{503}
\newcommand{\ffSysFnsWrappedPct}{95.3}  %
\newcommand{\ffSysFnsWrappedNotUsed}{15} %

\newcommand{\lgSysTysUsed}{73}
\newcommand{\lgSysTysWrapped}{68}
\newcommand{\lgSysTysWrappedPct}{93.2}  %
\newcommand{\ffSysTysUsed}{43}
\newcommand{\ffSysTysWrapped}{42}
\newcommand{\ffSysTysWrappedPct}{97.7}  %
\newcommand{\lgSysTysRaw}{5}  %
\newcommand{\ffSysTysRaw}{1}           %

\newcommand{\lgFields}{212}

\newcommand{\ffFields}{192}
\newcommand{\lgFieldAccRef}{213}
\newcommand{\lgFieldAccMut}{172}
\newcommand{\lgFieldAccOther}{8}
\newcommand{\ffFieldAccRef}{201}
\newcommand{\ffFieldAccMut}{217}
\newcommand{\ffFieldAccOther}{3}

\newcommand{\lgCtypeMacro}{40}
\newcommand{\lgCtypeManual}{34}  %

\newcommand{\ffCtypeMacro}{58}
\newcommand{\ffCtypeManual}{0}

\newcommand{\lgSigOwnerRet}{197}  %
\newcommand{\lgSigOwnerArg}{37}  %
\newcommand{\ffSigOwnerRet}{126}       %
\newcommand{\ffSigOwnerArg}{43}        %
\newcommand{\lgTyRawNoFfibox}{5}
\newcommand{\ffTyRawNoFfibox}{19}  %

\newcommand{\lgRefRest}{22}
\newcommand{\lgRefRestCmp}{16}

\newcommand{\ffRefRest}{25}
\newcommand{\ffRefRestCmp}{20}

\newcommand{\rhoTypeFields}{0.19}
\newcommand{\rhoTypeFieldsT}{0.41}
\newcommand{\medTypeNoFields}{5.19}
\newcommand{\medTypeFields}{6.58}
\newcommand{\usdPerType}{4.63}
\newcommand{\usdPerTypeMin}{0.77}
\newcommand{\usdPerTypeMax}{24.34}
\newcommand{\rhoSymLoc}{0.67}
\newcommand{\usdPerSym}{0.89}
\newcommand{\usdPerKloc}{52}
\newcommand{\usdPerTypeOpus}{4.95}
\newcommand{\usdPerKlocOpus}{58}
\newcommand{\usdPerTypeSol}{1.61}
\newcommand{\usdPerKlocSol}{11}
\newcommand{\lgReviewShare}{43}
\newcommand{\ffReviewShare}{52}
\newcommand{\fableReviewShare}{63}
\newcommand{\usdPerItemFable}{2.09}
\newcommand{\safePerHundredFable}{1.6}
\newcommand{\usdPerItemOpus}{0.87}
\newcommand{\safePerHundredOpus}{5.8}
\newcommand{\lgUsdPerItemBatchRv}{7.19}
\newcommand{\lgUsdPerItemWaveRv}{1.36}
\newcommand{\lgSafePerHundredBatchRv}{2.0}
\newcommand{\lgSafePerHundredWaveRv}{3.0}
\newcommand{\lgBatchRvCount}{4}
\newcommand{\lgSolTotItem}{1.96}
\newcommand{\lgOpusTotItem}{2.09}
\newcommand{\ffSolTotItem}{0.69}
\newcommand{\ffOpusTotItem}{3.13}

\newcommand{\lgWallHours}{51.9}
\newcommand{\ffWallHours}{58.9}

\newcommand{\scOurLines}{17{,}571}
\newcommand{\scOurFfimapUnsafeFn}{166}
\newcommand{\scOurFfimapUnsafeImpl}{324}
\newcommand{\scOurFfimapUnsafeBlock}{2{,}097}
\newcommand{\scOurFfimapUnsafeLines}{2{,}911}
\newcommand{\scOurFfimapFfiCall}{592}
\newcommand{\scOurFfimapPtrDeref}{443}
\newcommand{\scOurGeigerUnsafeFn}{156}
\newcommand{\scOurGeigerUnsafeImpl}{266}
\newcommand{\scOurGeigerUnsafeExpr}{6{,}448}
\newcommand{\scOurScanUnsafeImpl}{266}
\newcommand{\scOurScanFfiCall}{523}
\newcommand{\scOurScanEFfiCall}{540}
\newcommand{\scOurScanPtrDeref}{55}

\newcommand{\scOurClippyMultiOps}{0}
\newcommand{\scGitLines}{15{,}434}
\newcommand{\scGitFfimapUnsafeFn}{124}
\newcommand{\scGitFfimapUnsafeImpl}{8}
\newcommand{\scGitFfimapUnsafeBlock}{1{,}004}
\newcommand{\scGitFfimapUnsafeLines}{3{,}497}
\newcommand{\scGitFfimapFfiCall}{674}
\newcommand{\scGitFfimapPtrDeref}{291}
\newcommand{\scGitGeigerUnsafeFn}{123}
\newcommand{\scGitGeigerUnsafeImpl}{8}
\newcommand{\scGitGeigerUnsafeExpr}{3{,}336}
\newcommand{\scGitScanUnsafeImpl}{8}
\newcommand{\scGitScanFfiCall}{228}
\newcommand{\scGitScanEFfiCall}{646}
\newcommand{\scGitScanPtrDeref}{215}

\newcommand{\scGitClippyMultiOps}{400}
\newcommand{\scGeigerMissImpl}{58}
\newcommand{\scGeigerMissImplPct}{18}

\newcommand{\scOurMacroUnsafeFn}{18}

\newcommand{\scGeigerExprMorePct}{93}
\newcommand{\scBlocksRatio}{2.1}
\newcommand{\scLinesFewerPct}{17}
\newcommand{\scOurScanFfiPct}{88}
\newcommand{\scOurScanEFfiPct}{91}
\newcommand{\scGitScanEFfiDiffPct}{4}
\newcommand{\scOurFfimapCold}{37.2}
\newcommand{\scOurFfimapWarm}{12.9}
\newcommand{\scOurFfimapRss}{358}
\newcommand{\scOurGeigerCold}{12.2}
\newcommand{\scOurGeigerWarm}{11.4}
\newcommand{\scOurGeigerRss}{981}
\newcommand{\scOurScanCold}{38.2}
\newcommand{\scOurScanWarm}{26.8}
\newcommand{\scOurScanRss}{1{,}018}
\newcommand{\scOurScanECold}{46.2}
\newcommand{\scOurScanEWarm}{34.6}
\newcommand{\scOurScanERss}{1{,}017}
\newcommand{\scOurClippyCold}{9.8}
\newcommand{\scOurClippyWarm}{2.6}
\newcommand{\scOurClippyRss}{250}

\newcommand{\scOurFfimapRssShare}{0.36}
\newcommand{\scGitFfimapCold}{18.0}
\newcommand{\scGitFfimapWarm}{5.1}
\newcommand{\scGitFfimapRss}{245}
\newcommand{\scGitGeigerCold}{12.4}
\newcommand{\scGitGeigerWarm}{12.2}
\newcommand{\scGitGeigerRss}{1{,}530}
\newcommand{\scGitScanCold}{69.0}
\newcommand{\scGitScanWarm}{25.2}
\newcommand{\scGitScanRss}{1{,}181}
\newcommand{\scGitScanECold}{99.6}
\newcommand{\scGitScanEWarm}{55.9}
\newcommand{\scGitScanERss}{1{,}175}
\newcommand{\scGitClippyCold}{9.5}
\newcommand{\scGitClippyWarm}{0.45}
\newcommand{\scGitClippyRss}{25}
\newcommand{\scGitFfimapVsGeiger}{2.4}
\newcommand{\scGitFfimapRssShare}{0.16}
\newcommand{\scLoadMin}{19}
\newcommand{\scLoadMax}{41}

\renewcommand{\lgWallHours}{51.9}
\renewcommand{\ffWallHours}{58.9}

\newcommand{\lgPeakBatches}{32}

\newcommand{\lgSoloPct}{61}
\newcommand{\ffSoloPct}{58}
\newcommand{\lgAgentHoursSum}{112}
\newcommand{\ffAgentHoursSum}{159}
\newcommand{\lgExecWaves}{34}
\newcommand{\ffExecWaves}{26}
\newcommand{\lgWaveBatchesMedian}{2}
\newcommand{\ffWaveBatchesMedian}{2}
\newcommand{\lgWaveBatchesMax}{12}
\newcommand{\ffWaveBatchesMax}{15}
\newcommand{\lgMultiModWaves}{19}
\newcommand{\ffMultiModWaves}{14}
\newcommand{\lgSeqSpeedup}{2.2}  %
\newcommand{\ffSeqSpeedup}{2.7}  %

\newcommand{\lgTthreeRows}{119}
\newcommand{\ffTthreeRows}{126}
\newcommand{\lgTthreeAdd}{80}
\newcommand{\ffTthreeAdd}{91}
\newcommand{\lgTthreeDrop}{39}
\newcommand{\ffTthreeDrop}{35}
\newcommand{\ffTthreeErasedArg}{68}   %
\newcommand{\lgTthreeGenRows}{41}
\newcommand{\lgTthreeGenMacros}{4}
\newcommand{\lgTthreeFirst}{67}
\newcommand{\lgTthreeKept}{67}
\newcommand{\lgTthreeLater}{52}
\newcommand{\ffTthreeFirst}{69}
\newcommand{\ffTthreeKept}{68}
\newcommand{\ffTthreeLater}{58}

\begin{document}

\title{Crustify: Repository-Level Generation of Safe Rust Bindings with LLM Agents and Coverage-Guided Testing}

\author{Marius Momeu}
\affiliation{%
  \institution{University of California, Berkeley}
  \country{USA}}
\email{marius.momeu@berkeley.edu}
\orcid{0009-0009-3389-9837}

\author{Marius Arvinte}
\affiliation{%
  \institution{Intel Corporation}
  \country{USA}}
\email{marius.arvinte@intel.com}
\orcid{0000-0002-9328-9223}

\author{Cory Cornelius}
\affiliation{%
  \institution{Intel Corporation}
  \country{USA}}
\email{cory.cornelius@intel.com}
\orcid{0000-0002-6831-9910}

\author{Koushik Sen}
\affiliation{%
  \institution{University of California, Berkeley}
  \country{USA}}
\email{ksen@berkeley.edu}
\orcid{0000-0002-4539-9188}

\begin{abstract}
    Wrapping C code in a safe Rust interface lets Rust applications use it without writing \texttt{unsafe} code, until the code itself is migrated to Rust. Such bindings are tedious and error-prone to write, however: their internals use \texttt{unsafe} code that the Rust compiler does not check, so a single mistake can introduce \gls{ub}, and they have so far been written mostly by hand.
    
    We present Crustify, an agentic co-translator that generates safe Rust bindings for repository-level C libraries, with or without its user in the loop. Crustify schedules a target's types and symbols into dependency-ordered waves of independent batches that agents bind concurrently. Its agents derive the ownership and lifecycle contract of every pointer and type they wrap, check each binding with coverage-guided equivalence and \gls{ub} tests under sanitizers and Miri, and review every wave before its dependents start. They build on three tools that we contribute: (a) \texttt{ffibox}, a framework of Rust primitives for C/Rust interop that mirror \texttt{Box}, \texttt{Arc}, and \texttt{Vec} for C objects without ever forming a reference to memory that C may alias; (b) Wavefront, a neuro-symbolic static analyzer that orders a target's items by their dependencies, including those that casts and type-erased pointers hide; and (c) ffimap, a rustc-based analysis that maps the \texttt{unsafe} surface that bindings expose.
    
    On libgit2 and ffmpeg, with one developer in the loop, Crustify generated safe bindings for \lgApiCov\% and \ffApiCov\% of their public APIs in about \lgWallHours{} and \ffWallHours{} hours, respectively. On the interface that consumers are expected to use, only \lgUnsafeFnOtherPct\% and \ffUnsafeFnOtherPct\% of the bindings' public functions are \texttt{unsafe}, and only \lgRawPtrFnsPct\% and \ffRawPtrFnsPct\% have a raw pointer in their signature. Their tests cover \lgCovRust\% and \ffCovRust\% of the Rust code and \lgCovC\% and \ffCovC\% of libgit2 and libavutil, the only ffmpeg library wrapped so far, and the reviewers fixed \lgReportsUb{} and \ffReportsUb{} defects that caused \gls{ub} reachable from Rust. To our knowledge, this is the first work to automate the generation of safe Rust bindings for C libraries.

\end{abstract}


\maketitle
\glsresetall

\section{Introduction}
\label{sec:intro}

Translating C code to safe, production-ready Rust is time-consuming and costly. Until a library is translated, Rust applications can only use it through \texttt{unsafe} \gls{ffi} calls and raw pointers, and thus lose Rust's memory-safety guarantees. An intermediate step is to \emph{wrap} the C code in a \emph{safe Rust binding} whose public interface is safe, while its implementation calls the C code through the \gls{ffi} in \texttt{unsafe} blocks. Rust's standard library follows the same design: types such as \texttt{Cell}, \texttt{RefCell}, \texttt{Mutex}, and \texttt{Arc} expose a safe interface over an implementation that uses \texttt{unsafe} code~\cite{jung2018rustbelt}. A safe binding should be both \emph{sound}, i.e., no program that uses it without writing \texttt{unsafe} code can cause \gls{ub}, and \emph{ergonomic}, i.e., its consumers need not uphold any memory-safety discipline themselves: they write no \texttt{unsafe} code, handle no raw pointers, release no C objects by hand, and need not remember to avoid forming references to memory that C may alias, but use the idioms they know from Rust's standard library. An ergonomic binding is thus also safe to use, since its consumers write no \texttt{unsafe} code; we call the artifacts \emph{safe bindings} throughout, and use \emph{ergonomic} for the property that their design targets. Designing such an interface is non-trivial, however, since the Rust compiler does not verify its soundness---the implementer must ensure it.

Safe bindings have so far been written mostly by hand. Automated tools either emit raw, \texttt{unsafe} bindings, e.g., bindgen~\cite{bindgen}, or translate C code to Rust instead of generating safe bindings for it, e.g., C2Rust and its successors~\cite{c2rust,emre2021laertes,zhang2023crown} and translators based on \glspl{llm}~\cite{nitin2026c2saferrust,hong2025tymcrat,smartc2rust2026,rustmap2025,ptrtrans2026,evoc2rust2026,sactor2026,cornelius2026ideas,recodeagent2026,schesch2026agentic}. Existing open-source projects that publish safe bindings over widely used C libraries took years to develop and still lack complete coverage of the original API~\cite{git2rs,ffmpegnext,ffmpegthethird,rustopenssl,rusqlite,curlrust}. Moreover, no widely adopted framework of primitives exists for representing types and pointers that cross the \gls{ffi} boundary in safe Rust, so most projects write their own variants. Such hand-written encapsulations are prone to \gls{ub}: McCormack et al.~\cite{mccormack2025ffi} found 46 instances of undefined or undesired behavior in 37 Rust libraries that call foreign functions.

In this paper, we present \emph{Crustify}, an agentic co-translator that generates safe Rust bindings over C APIs, designed to be sound and ergonomic, and scales to repository-level targets. Crustify consists of three tools, which we introduce below, and a common, human-reviewed specification that its \gls{llm} agents follow. An \emph{assistant} agent establishes the campaign's scope and settings with the user and drives the campaign through the agents it spawns: a \emph{planner} decomposes the target into \emph{modules} and schedules their types and symbols into dependency-ordered waves of size-bounded batches; \emph{binders} bind and test each batch; \emph{reviewers} audit each landed wave for \gls{ub}, equivalence divergences from the C implementation, and coverage gaps, and fix the defects they find. The batches of a wave are independent by construction, so Crustify processes them concurrently as isolated agent sessions in separate worktrees. Since every agent's prompt is assembled from the specification, independent sessions apply the same rules, and a rule changed once changes every agent that follows it (\S\ref{sec:agents}). To our knowledge, Crustify is the first agentic system that automates the generation of safe Rust bindings for C libraries.

We also develop \texttt{ffibox}, a Rust framework of generic primitives that represent C types and pointers that cross the \gls{ffi} boundary, built around one invariant: no Rust reference to memory that C may alias is ever formed. Policy-typed owners that mirror \texttt{Box}, \texttt{Arc}, and \texttt{Vec} are bound to the C routines that allocate, clone, and free each object, and lifetime-annotated handles encapsulate raw-pointer accesses without forming \texttt{\&} or \texttt{\&mut}, so that consumers use the idioms of native Rust while the C-specific parts stay internal to the binding. This lets a binding express owned, shared, borrowed, and lock-protected C objects, as well as length-bounded buffers and NUL-terminated strings, while ruling out the aliasing violations that references into C memory cause (\S\ref{sec:ffibox}). To our knowledge, \texttt{ffibox} is the first framework of Rust primitives that expresses such pointer-level properties without references into C memory.

Wrapping a repository-level target also depends on ordering: a binding can use the safe wrappers of the types it depends on only if they are wrapped first, and otherwise falls back to their raw \gls{ffi} types and thus to \texttt{unsafe} code. C hides many of these dependencies, since programs emulate polymorphism with casts between struct types and pass objects through type-erased \texttt{void *} slots, and large repositories form dependency cycles. Prior work recovers such casts to check or retype C code~\cite{siff1999casts,wu2025genc2rust}, whereas project-level C-to-Rust translators order their units by calls and type references only, and merge or reject cycles, or break them wherever their traversal does (\S\ref{sec:related}). We therefore develop \emph{Wavefront}, a neuro-symbolic static analyzer that extracts with CodeQL the relations the code states unambiguously, such as field and signature types, calls, and direct casts between structs, and lets agents add those that require judgment, such as a \texttt{void *} that passes through several calls before it is cast, in correction tables that it checks for well-formedness. Wavefront performs no data-flow analysis; agents' judgment covers the relations that data flow would recover. It also breaks the graph's cycles and schedules it into \emph{waves} of mutually independent, size-bounded batches, which Crustify runs sequentially within each module, while independent modules and the batches of each wave run concurrently (\S\ref{sec:wavefront}). To our knowledge, Wavefront is the first to order C-to-Rust work by the relations that casts and type-erased slots hide.

Moreover, a binding depends on the ownership and lifecycle contract of every pointer and type that it exposes, i.e., whether a pointer owns or borrows its pointee and which routines construct, release, and clone each type. Prior work infers subsets of these properties, e.g., the ownership of pointers~\cite{zhang2023crown,ptrtrans2026,chen2026inator}, their bounds~\cite{mohammed2025msa}, or allocation and deallocation functions~\cite{shemetova2025lamed}. In Crustify, agents derive these contracts for every pointer and type they bind, store them in Wavefront's \emph{ownership store} for later batches and reviewers (\S\ref{sec:wavefront}), and validate them through coverage-guided tests under sanitizers (\S\ref{sec:testing}) and through review (\S\ref{sec:review}). To our knowledge, Crustify is the first system in which \gls{llm} agents derive both the ownership and the lifecycle contracts of a C library, including dual ownership, reference-counted cloning, and lock-protected state.

Measuring how far a binding falls short of these goals also lacks tooling: existing tools count \texttt{unsafe} code syntactically or by its declared visibility, rather than by what a crate's consumers can reach, and those that parse unexpanded source miss the \texttt{unsafe} items that macros generate (Appendix~\ref{sec:appendix-ffimap}). We therefore develop \emph{ffimap}, a rustc driver that maps a crate's \gls{ffi} surface after type checking. Its generic counters, which apply to any crate, locate the public \texttt{unsafe} functions and the raw pointers on the safe public interface; its \texttt{ffibox} counters measure how much of the C surface a binding wraps with \texttt{ffibox}, and where it bypasses it (\S\ref{sec:scan}). Agents use ffimap as a linter, and we use it to measure how consistently they built on \texttt{ffibox} rather than reinventing their own primitives, and to compare our bindings with community-maintained ones.

Checking the soundness of safe bindings currently lacks a systematic workflow. The three community-maintained bindings that we compare against run their test suites and lints in \gls{ci}, but no sanitizers, Miri~\cite{miri2026}, or coverage measurement. Crustify makes these checks part of every batch. Its binders emit equivalence tests that call both the original C function and its Rust wrapper, and guide test generation by line coverage on both sides of the boundary, extending coverage-guided \gls{llm} test generation~\cite{codamosa2023,coverup2025} to cross-language targets, whereas C-to-Rust harnesses mostly run existing test suites against the translation~\cite{smartc2rust2026,rustmap2025,sactor2026,nitin2026c2saferrust}, or generate equivalence tests without measuring coverage across the boundary~\cite{cornelius2026ideas,schesch2026agentic}, and report coverage only as an evaluation metric~\cite{rustassure2025,ptrtrans2026,sactor2026}. Crustify runs these tests under Miri, to detect \gls{ub} in unsafe Rust, and under AddressSanitizer, UndefinedBehaviorSanitizer, ThreadSanitizer, and BorrowSanitizer~\cite{serebryany2012asan,tsan2009,serebryany2011tsanllvm,ubsan,mccormack2025bsan}, to detect memory errors, undefined behavior, data races, and aliasing violations across the boundary---a combination that, to our knowledge, no prior C-to-Rust work applies to translated code.

We evaluated Crustify on libgit2 and ffmpeg, whose public APIs comprise \lgApiTypes{} and \ffApiTypes{} types, and \lgApiSyms{} and \ffApiSyms{} symbols, mainly with \texttt{claude-opus-5.5}. In about \lgWallHours{} and \ffWallHours{} hours of wall-clock time and about \lgTokensB{} and \ffTokensB{} billion tokens (\$\lgCostUSD{} and \$\ffCostUSD{}), with one developer checking in between its sub-campaigns, Crustify wrapped \lgApiCov\% and \ffApiCov\% of the two APIs, with tests that cover \lgCovRust\% and \ffCovRust\% of the Rust bindings and \lgCovC\% and \ffCovC\% of libgit2 and libavutil, the only ffmpeg library wrapped so far, against \gitCovRust\% and \gitCovC\% for git2-rs (\lgCovCCmp\% for ours over the same C, i.e., libgit2 with its bundled dependencies) and at most \nextCovRust\% and \nextCovC\% for the community-maintained ffmpeg bindings (\S\ref{sec:eval-comparison}). On the interface that consumers are expected to use, i.e., excluding the seams that convert between wrappers and C values and the methods that \texttt{ffibox}'s traits declare \texttt{unsafe}, only \lgUnsafeFnOtherPct\% and \ffUnsafeFnOtherPct\% of the bindings' public functions are \texttt{unsafe}, and only \lgRawPtrFnsPct\% and \ffRawPtrFnsPct\% have a raw pointer in their signature. As ffimap measures, the bindings build on \texttt{ffibox} rather than on primitives of their own: \lgTyFfiboxPct\% and \ffTyFfiboxPct\% of their public types implement its traits, as layout types and their handles or as lifecycle policies, and \lgSysTysWrappedPct\% and \ffSysTysWrappedPct\% of the C structs they use behind pointers have such a layout type (\S\ref{sec:eval-guarantees}). The review agents found and fixed \lgReportsUb{} and \ffReportsUb{} defects that caused \gls{ub} reachable from Rust, each confirmed by a reproducer.

In summary, we make the following contributions:
\begin{itemize}
  \item \emph{Crustify}, a spec-driven agentic co-translator that generates safe Rust bindings for repository-level C libraries (\S\ref{sec:crustify});
  \item \texttt{ffibox}, Rust primitives for C types and pointers that never form a reference to memory that C may alias (\S\ref{sec:ffibox});
  \item \emph{Wavefront}, a neuro-symbolic analyzer that schedules a C target into waves of independent batches and stores agents' ownership contracts (\S\ref{sec:wavefront});
  \item \emph{ffimap}, a rustc-based analysis of a crate's \texttt{unsafe} surface and its use of \texttt{ffibox} (\S\ref{sec:scan}); and
  \item safe bindings for libgit2 and ffmpeg, covering \lgApiCov\% and \ffApiCov\% of their public APIs (\S\ref{sec:eval}).
\end{itemize}

We release the Crustify harness, \texttt{ffibox}, Wavefront, ffimap, and the generated safe bindings for libgit2 and ffmpeg as open source at \url{https://github.com/crustify-rs}.

\section{Background}
\label{sec:background}

\subsection{Rust Safety and Soundness}
\label{sec:bg-rust}

\paragraph{Ownership, borrowing, and thread safety.}
Rust guarantees memory safety without garbage collection by checking ownership and borrowing at compile time: every value has a unique owner, which releases it through its type's \texttt{Drop} implementation, e.g., \texttt{Box<T>} is the sole owner of a heap allocation and frees it when dropped; code that does not own a value accesses it through a shared reference \texttt{\&T} or a mutable reference \texttt{\&mut T}, whose \emph{lifetime} the borrow checker bounds by that of its referent. References obey \emph{aliasing XOR mutability}: a value is accessed either through any number of shared references or through exactly one mutable reference, unless it is wrapped in an \texttt{UnsafeCell<T>}, which permits \emph{interior mutability}, as in \texttt{Mutex<T>} and \texttt{RefCell<T>}, which enforce exclusivity at run time. The compiler relies on this rule to optimize, so a program that violates it, e.g., by mutating memory through a C-held pointer while a reference to it is live, has \gls{ub}. Stacked Borrows~\cite{jung2020stackedborrows} and its successor Tree Borrows~\cite{villani2025treeborrows} define which access patterns are \gls{ub}. A type is \texttt{Send} if its values can be moved to another thread, and \texttt{Sync} if they can be shared between threads. \texttt{Arc<T>}, which shares a value through a reference count that \texttt{Clone} increments and \texttt{Drop} decrements, is \texttt{Send} and \texttt{Sync} only when \texttt{T} is both, and hands out a \texttt{\&mut T} only through \texttt{Arc::get\_mut}, when the reference count proves it the sole owner.

\paragraph{Unsafe code and soundness.}
The compiler cannot verify portions of programs contained in \texttt{unsafe} blocks. These blocks may dereference raw pointers (\texttt{*const T}, \texttt{*mut T}), call \texttt{unsafe} functions, including through an \gls{ffi}, or implement \texttt{unsafe} traits such as \texttt{Send} and \texttt{Sync}. Raw pointers carry no ownership or lifetimes, so the obligation to uphold Rust's invariants is left to the programmer. Libraries therefore encapsulate \texttt{unsafe} code behind safe APIs, whose implementation is \emph{sound}, as defined in \S\ref{sec:intro}, if no program that uses the API without writing \texttt{unsafe} code can cause \gls{ub}~\cite{jung2018rustbelt}. Safe bindings over C libraries are an instance of this design pattern: their implementation calls foreign functions and handles raw pointers, while their API is marked as safe and must uphold Rust's ownership, aliasing, and thread-safety guarantees although the borrow checker does not check the foreign code. By convention, a C library's raw declarations, which bindgen~\cite{bindgen} generates from its headers, form a \texttt{-sys} crate, over which the safe crate is written.

Rust's smart pointers, however, are ill-suited for objects shared with C. For example, a \texttt{Box<T>} can adopt a C-allocated object through its \texttt{unsafe} \texttt{from\_raw} constructor, but it will release the object using Rust's allocator rather than the library's possibly customized destructor, and it hands out \texttt{\&T} and \texttt{\&mut T} references to the object, whose aliasing guarantees the compiler assumes but cannot enforce, since the C side may retain and use its own pointers to it. A binding cannot simply replace these references with lifetime-bound handles: the dereference traits \texttt{Deref} and \texttt{DerefMut} must return a plain reference, and the standard library already implements them for every \texttt{Box<T>}. Similarly, \texttt{Arc<T>} keeps its reference count in a header that it allocates itself, whereas many C objects maintain their own count, incremented and decremented through customized library routines. \texttt{ffibox} addresses these limitations (\S\ref{sec:ffibox}).

\subsection{Sanitizers and Miri}
\label{sec:bg-sanitizers}

Sanitizers instrument a program at compile time and report an error when an execution violates the invariants they check: \gls{asan}~\cite{serebryany2012asan} detects temporal and spatial memory errors, \gls{ubsan}~\cite{ubsan} operations that are undefined in C, such as signed integer overflow, and \gls{tsan}~\cite{tsan2009,serebryany2011tsanllvm} data races. \gls{msan}~\cite{msan2015} detects reads of uninitialized memory, but requires the whole program, including its libraries, to be instrumented. Sanitizers instrument both Rust and C code, and so detect errors across the \gls{ffi} boundary, but only on the paths that tests execute, and none of them checks Rust's aliasing rules. Two tools check these rules. Miri~\cite{miri2026}, an interpreter for the Rust compiler's \gls{mir}, detects \gls{ub} in Rust code, including violations of Stacked or Tree Borrows, but runs orders of magnitude slower than native code and does not check the foreign code that it calls. BorrowSanitizer~\cite{mccormack2025bsan}, which is still under development, instruments Rust together with statically linked C code to detect violations of Tree Borrows, including those caused by accesses on the C side. Crustify's agents use these tools to catch regressions and to reproduce the flaws they find.

\subsection{LLM Agents}
\label{sec:bg-agents}

An \gls{llm} agent runs a language model in a loop that interleaves reasoning with actions, such as tool calls, whose results inform its next step~\cite{yao2023react}. Such agents resolve real GitHub issues by navigating a repository, editing its files, and running its tests~\cite{yang2024sweagent,zhang2024autocoderover}. Coding agents such as Claude Code~\cite{claudecode} and Codex~\cite{codex} run this loop in the user's workspace with tools to search, read, and edit files and to execute shell commands; Crustify runs each of its agents as a session of such an application (\S\ref{sec:agents}). When a session approaches the model's context limit, the application \emph{compacts} it into a summary written by the model, which may omit instructions given early in the session, so persistent instructions are kept in files that the application loads into every session~\cite{claudecode}.

A model's accuracy declines as its input grows~\cite{liu2024lost,hsieh2024ruler,du2025context} and as the number of steps in a task increases~\cite{sinha2026illusion}, which decomposing a task into simpler subtasks can mitigate~\cite{khot2023decomposed}; Crustify therefore assigns each agent a single batch of bounded size (\S\ref{sec:wavefront}). Without external feedback, models do not improve their answers to reasoning tasks by revising them, and at times change correct answers into incorrect ones~\cite{huang2024cannot}, whereas feedback from executing the code~\cite{chen2024selfdebug} or from a stronger model~\cite{olausson2024selfrepair} helps, and \gls{llm} evaluators favor their own outputs~\cite{panickssery2024selfpref}. Crustify's agents therefore receive external feedback from the compiler and linter, from \gls{ub}, equivalence, and unit tests run under sanitizers and Miri, from line coverage on both sides of the \gls{ffi} boundary, and from ffimap, and its reviewers can run on a different model than the binders and accept a \gls{ub} or equivalence finding only if it reproduces through the Rust API (\S\ref{sec:review}).

\section{Crustify}
\label{sec:crustify}

Given a C repository, Crustify generates a safe Rust crate that wraps its public API and is designed to be both sound and ergonomic (\S\ref{sec:intro}): rather than uphold a memory-safety discipline themselves, its consumers use the idioms that Rust programmers know from the standard library, i.e., owners that release on \texttt{Drop} and duplicate on \texttt{Clone}, borrows that the borrow checker keeps from outliving what they reference, and errors as \texttt{Result}s. We therefore design for ergonomics by mirroring the standard library's primitives as closely as C allows, e.g., with owners that behave as a \texttt{Box} or an \texttt{Arc}, and borrowed arrays that behave as slices. The crate confines \texttt{unsafe} code to its internals, behind the safe API, and Crustify checks it with coverage-guided tests under sanitizers and Miri and with agentic review. 

\begin{figure}
  \centering
  \begin{tikzpicture}[
    x=1mm, y=1mm,
    font=\footnotesize,
    >={Stealth[length=1.6mm]},
    box/.style={draw, rounded corners=1pt, align=center, inner sep=1.5pt, fill=white},
    stepbox/.style={box, text width=18mm, minimum height=7mm},
    artifact/.style={font=\scriptsize\ttfamily, text=black!65},
    doc/.style={box, fill=yellow!10},
    agent/.style={box, fill=blue!8, minimum height=7mm, text width=11.5mm},
    rev/.style={box, fill=orange!12, minimum height=7mm, text width=11.5mm},
    tool/.style={box, fill=black!4, minimum height=7mm, text width=18mm},
    phase/.style={draw=black!40, rounded corners=2pt, inner sep=2mm},
    plabel/.style={font=\footnotesize\bfseries, anchor=south west, inner sep=1pt},
    sub/.style={font=\scriptsize},
    flow/.style={->, thick},
    talk/.style={->, thick, green!40!black},
    use/.style={<->, >={Stealth[length=1.1mm]}, densely dashed, thick, black!55},
    feed/.style={->, densely dotted, thick, black!55},
  ]

  \node[box, text width=15mm, minimum height=8mm] (user) at (8.5,-4) {User};
  \node[doc, text width=16mm, minimum height=8mm] (settings) at (8.5,-17)
    {\texttt{CRUSTIFY.md}\\[-1pt]\scriptsize settings};
  \node[box, text width=15mm, minimum height=8mm] (repo) at (8.5,-30) {C repository};
  \node[box, fill=green!4, text width=15mm, minimum height=8mm] (crate) at (8.5,-43)
    {Safe Rust crate};

  \node[box, fill=green!8, text width=14mm, minimum height=53mm, anchor=north] (asst) at (32,0.5)
    {\textbf{Assistant}\\[-1pt]\scriptsize co-translator};

  \draw[talk] ([yshift=1.6mm]user.east) -- node[sub, above, inner sep=1pt] {goals} ([yshift=1.6mm]user.east -| asst.west);
  \draw[talk] ([yshift=-1.6mm]user.east -| asst.west) -- node[sub, below, inner sep=1pt] {brief} ([yshift=-1.6mm]user.east);
  \draw[flow] (settings.east) -- (settings.east -| asst.west);
  \draw[flow] (repo.east) -- (repo.east -| asst.west);
  \draw[flow] (crate.east -| asst.west) -- (crate.east);

  \coordinate (colL) at (45.5,0);
  \coordinate (colR) at (109,0);
  \node[agent, fill=violet!10, text width=10mm] (planner) at (54,-7) {Planner};
  \node[stepbox] (decomp) at (74.5,-7) {Module\\[-1pt]decomposition};
  \node[stepbox] (sched)  at (98.5,-7) {Scheduling};
  \draw[flow] (planner) -- (decomp);
  \draw[flow] (decomp) -- (sched);
  \node[artifact, below=0.3mm of decomp, align=center] (a2) {modules.json\\[-2pt]dependencies.json};
  \node[artifact, below=0.3mm of sched]  (a3) {schedule.json};

  \begin{scope}[on background layer]
    \node[phase, fit=(planner)(sched)(a2)(planner -| colL)(planner -| colR)] (planning) {};
  \end{scope}
  \node[plabel] at (planning.north west) {Planning};

  \node[box, text width=13.5mm, minimum height=4.4mm, sub, draw=black!45, text=black!65] (m0) at (52.5,-27)
    {foundational};
  \node[box, text width=13.5mm, minimum height=4.4mm, sub, fill=blue!8, very thick] (m1) at (52.5,-33)
    {module $j$};
  \node[sub] (md) at (52.5,-39.5) {$\vdots$};
  \node[box, text width=13.5mm, minimum height=4.4mm, sub, draw=black!45, text=black!65] (mn) at (52.5,-46)
    {module $n$};
  \draw[->, black!45] (m0) -- (m1);

  \node[agent] (b1) at (71,-32) {Binder\\[-1pt]\scriptsize batch 1};
  \node        (bd) at (80,-32) {$\cdots$};
  \node[agent] (bk) at (89,-32) {Binder\\[-1pt]\scriptsize batch $k$};
  \node[rev]   (r1) at (71,-44.5) {Reviewer\\[-1pt]\scriptsize batch 1};
  \node        (rd) at (80,-44.5) {$\cdots$};
  \node[rev]   (rm) at (89,-44.5) {Reviewer\\[-1pt]\scriptsize batch $m$};

  \begin{scope}[on background layer]
    \node[draw=blue!40, fill=blue!3, rounded corners=1pt, inner sep=1.2mm, fit=(b1)(bk)] (wave) {};
    \node[draw=orange!50, fill=orange!4, rounded corners=1pt, inner sep=1.2mm, fit=(r1)(rm)] (review) {};
  \end{scope}
  \node[sub, anchor=west] (wl) at ([xshift=0.3mm]wave.east) {wave $i$};
  \node[sub, anchor=west, align=left] (rl) at ([xshift=0.3mm]review.east) {review\\[-2pt]wave $i$};

  \draw[flow] (m1.east) -- (m1.east -| wave.west);
  \draw[flow] (wave.south) -- (review.north);

  \begin{scope}[on background layer]
    \node[phase, fit=(m0)(mn)(wave)(review)(wl)(rl)(m1 -| colL)(m1 -| colR)] (waves) {};
  \end{scope}
  \node[plabel] at (waves.north west) {Binding and review waves};

  \draw[flow] (a3.south) -- (a3.south |- waves.north);
  \draw[flow] (asst.east |- planner) -- (planning.west |- planner);
  \draw[flow] (asst.east |- m1) -- (waves.west |- m1);

  \node[tool] (t1) at (127.8,-6)  {\texttt{ffibox}\\[-1pt]\scriptsize primitives};
  \node[tool] (t2) at (127.8,-18) {Wavefront\\[-1pt]\scriptsize graph, ownership};
  \node[tool] (t3) at (127.8,-30) {ffimap\\[-1pt]\scriptsize safety scans};
  \node[tool] (t4) at (127.8,-42) {Tests\\[-1pt]\scriptsize sanitizers, Miri};
  \begin{scope}[on background layer]
    \node[phase, draw=black!25, fill=black!2, fit=(t1)(t4), inner sep=1.5mm] (tools) {};
  \end{scope}
  \node[plabel] at (tools.north west) {Agent tools};

  \draw[use] (planning.east) -- (tools.west |- planning.east);
  \draw[use] (waves.east) -- (tools.west |- waves.east);

  \node[doc, text width=110mm, minimum height=5.5mm, anchor=north, font=\scriptsize] (spec) at (80,-57.5)
    {\textbf{Specification}: workflows, decomposition, scheduling, binding, testing, review, conventions};
  \draw[feed] (spec.north -| asst) -- (asst.south);
  \draw[feed] (spec.north -| b1) -- (waves.south -| b1);
  \draw[feed, rounded corners=2pt] ($(spec.north -| waves.east)!0.5!(spec.north -| tools.west)$) |- ([yshift=-4.5mm]planning.east);
\end{tikzpicture}
  \caption{Overview of Crustify. The assistant scopes the campaign with the user; a planner decomposes the target into modules and schedules their items into waves; binders and reviewers process each wave's batches concurrently; all agents follow a common specification and use the same tools.}
  \Description{Block diagram: the user exchanges goals and a brief with the assistant, which reads the CRUSTIFY.md settings and drives a planning stage (planner, module decomposition, scheduling) and per-module binding and review waves of concurrent binder and reviewer batches over the C repository, producing a safe Rust crate. All agents follow a shared specification and use ffibox, Wavefront, ffimap and the test tools.}
  \label{fig:architecture}
\end{figure}
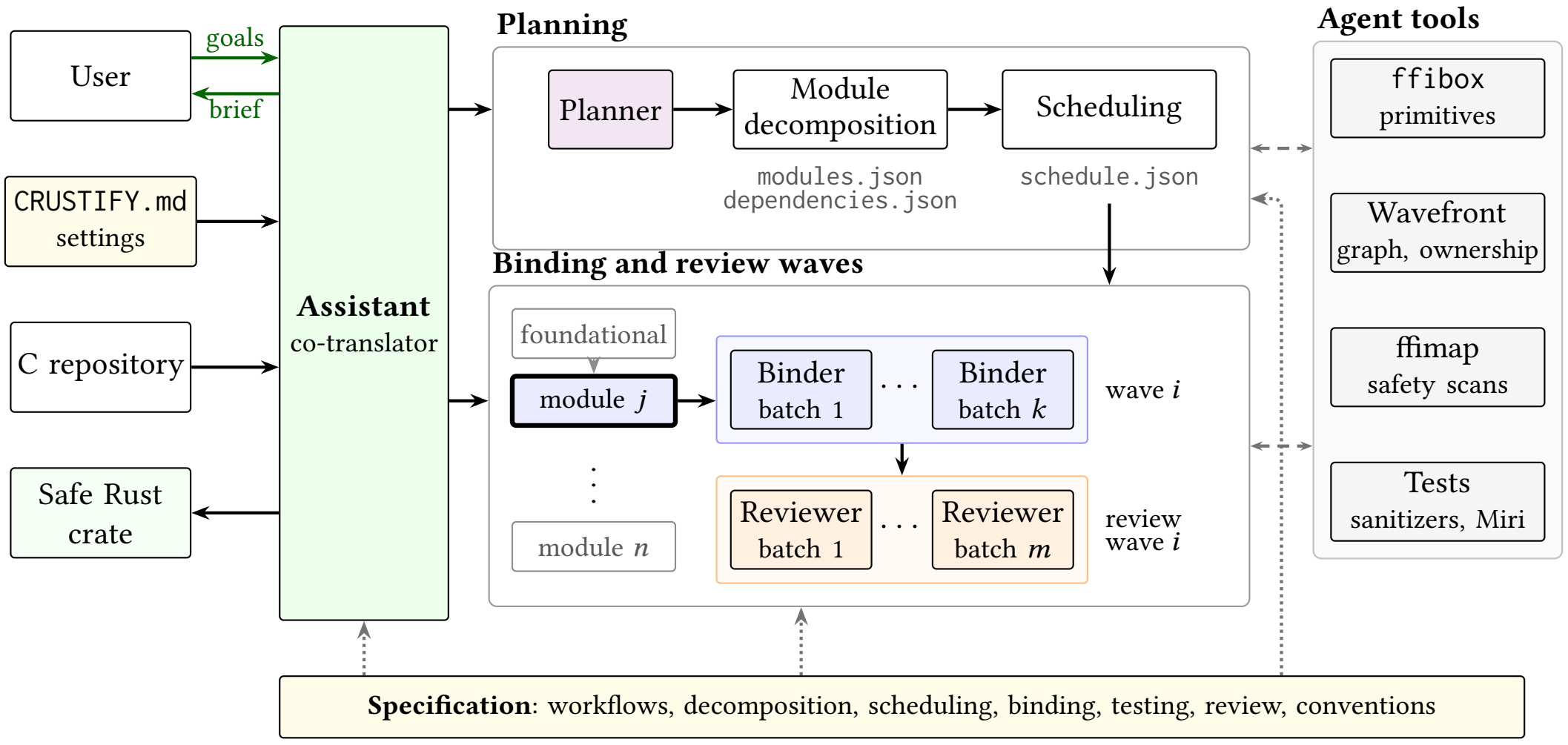

Crustify divides the work between \gls{llm} agents and deterministic tools: agents run tools, generate code and tests, and make the semantic judgments that deterministic tools cannot, while the tools take on the repeatable, mechanical tasks. It consists of three tools and a common specification that \gls{llm} agents follow (Figure~\ref{fig:architecture}, \S\ref{sec:crustify-workflow}): \texttt{ffibox} provides the C/Rust interop primitives from which agents build the bindings (\S\ref{sec:ffibox}), ffimap maps their \texttt{unsafe} surface and their use of \texttt{ffibox} primitives (\S\ref{sec:scan}), and Wavefront schedules the target's types and symbols into ordered waves of independent batches, answers the agents' queries about the code base, and stores the ownership and lifecycle contracts they derive (\S\ref{sec:wavefront}). An assistant agent, Crustify's co-translator, drives the campaign through three kinds of agents that it spawns: \emph{planners} decompose the target into modules and schedule their types and symbols into waves of batches, a \emph{binder} wraps each batch of a wave, and a \emph{reviewer} audits each batch of the review wave that follows it (\S\ref{sec:crustify-workflow}, \S\ref{sec:agents}). The user fills in the campaign's settings, live or in \texttt{CRUSTIFY.md}, approves its brief, and can intervene at any point; between such approval gates, the assistant runs autonomously.

\subsection{ffibox: Primitives for Safe FFI Bindings}
\label{sec:ffibox}

A designer of a safe binding interface must decide how every C type and pointer that crosses the \gls{ffi} boundary is owned, borrowed, shared, and released---and must do so without introducing \gls{ub}, since the Rust compiler cannot check the \texttt{unsafe} code that implements these decisions. This requires recovering from the C code each type's lifecycle routines, i.e., its constructors, destructors, and cloners, and each pointer's properties: the type it points to, whether it owns or borrows its pointee, whether access through it is shared or exclusive, whether it may be null, and whether it points to a single object or to an array. C often settles these properties only at run time or by convention: ownership may depend on a flag (\S\ref{sec:synthesis}), pointers may be type-erased (\texttt{void *}), a type may have several destructors or separate its allocation from its initialization, and a pointer to an array looks like a pointer to a single object.

\texttt{ffibox} encodes these decisions in a small set of generic primitives that mirror Rust's standard ones (Table~\ref{tab:ffibox}), so that Rust's type and borrow checkers enforce them. It thus serves both goals of a safe crate (\S\ref{sec:intro}). For ergonomics, its primitives give consumers Rust's ownership, borrowing, and threading idioms over C objects through the interfaces of their standard counterparts, e.g., a \texttt{CBox} releases its object when dropped and a \texttt{CArc} shares it when cloned, as \texttt{Box} and \texttt{Arc} do. For soundness, their \texttt{unsafe} code is written and checked once, in \texttt{ffibox}, and what each binding must still justify is a fixed set of \texttt{unsafe} trait implementations, one per type and obligation (C6), most of which macros emit, so that auditing a binding's soundness reduces to checking these implementations against the C routines they name, rather than every function that touches C. The primitives are built around one invariant, which rules out the aliasing violations that references into C memory cause (\S\ref{sec:bg-rust}): \emph{no Rust reference (\textup{\texttt{\&}} or \textup{\texttt{\&mut}}) to memory that C may alias is ever formed}. We describe the primitives along six challenges (C1--C6). The invariant has one exception: a global that is immutable after initialization is exposed as a plain shared reference, sound because neither side writes it (\S\ref{sec:synthesis}).

\paragraph{(C1) Type layout.}
Each C type \texttt{foo} gets a layout type \texttt{Foo}, a \texttt{\#[repr(transparent)]} newtype over the bindgen type with private fields and no \texttt{Drop}. Code outside the binding crate thus cannot build a \texttt{Foo} from a raw value, and releasing it is left to its owner (C2). A zero-sized \texttt{PhantomData<*const ()>} field withholds \texttt{Send} and \texttt{Sync}, requiring types to opt in explicitly (C4). Field accessors live on the borrowed handles (C3), never on \texttt{Foo}, so the only references that cover a C struct's bytes are those to a \texttt{Foo} that Rust holds inline, e.g., a zero-initialized local or the contents of a \texttt{CVal} (C2). Such a value moves by copy and is therefore sound only while C holds no pointer to it, so the binding decides which types consumers may hold by value: a type that C registers or that is address-sensitive (e.g., self-referential) is held only behind a \texttt{CBox}.

\paragraph{(C2) Ownership across the boundary.}
\texttt{ffibox} provides three owners: \texttt{CBox<Foo, P>} is the sole owner of an object behind a pointer and releases it through the policy \texttt{P}, mirroring \texttt{Box}; \texttt{CArc<Foo, P>} is one of several reference-counted owners, mirroring \texttt{Arc}; and \texttt{CVal<Foo, P>} holds an object inline and disposes of its resources (e.g., \texttt{*\_clear}) on drop. The \emph{policy} \texttt{P} names the C routine that releases the object, and the routines that allocate and clone it where they exist; it is usually a zero-sized type, but may carry run-time state as fields for routines that need data the object does not store. Since these routines are bound to the policy rather than to \texttt{Foo}, a type may have several policies, i.e., several pairs of constructors and destructors. A policy that also allocates, e.g., by calling the type's C constructor or a generic memory allocator, gives \texttt{CBox} and \texttt{CArc} safe constructors, so an object built in safe Rust is always released by the routine paired with its allocator.

\paragraph{(C3) Borrowing C memory.}
Forming a native \texttt{\&mut Foo} over memory that C may still alias would assert uniqueness that C does not respect, which is \gls{ub}. Such a reference would also assert that the bytes are valid, which holds only once C has initialized them, often in stages after allocating the storage. Instead, every layout type defines two lifetime-annotated handles, \texttt{FooRef<'a>} and \texttt{FooMut<'a>}, each one pointer wide and bound to the memory it references. The getters and setters that Crustify synthesizes live on the handles and project fields through raw pointers, so the borrow checker tracks the handles' lifetimes while no reference into C memory is created. The owners of C2 hand out only these handles, never a native reference to the object.

\paragraph{(C4) Sharing across threads.}
Wrapped types withhold \texttt{Send} and \texttt{Sync} by default; a wrapper opts in with an explicit safety argument. \texttt{CArc} hands out the exclusive handle only when the reference count proves it the sole owner, and for objects protected by a C lock, a \texttt{CGuard} exposes a \texttt{FooLocked<'a>} handle that reaches the protected state only while the lock is held. Both rely on traits that register the locking routines and the ownership check for a given type (C6).

\paragraph{(C5) Buffers and strings.}
\texttt{CVec<T, P>} owns a \texttt{(ptr, len)} array released by a C destructor, mirroring \texttt{Vec}, and \texttt{CStrBox<P>} owns a NUL-terminated \texttt{char *} with read-only \texttt{CStr}/\texttt{str} views. Borrowed runs of wrapped C objects---a buffer's, a struct field's, or a getter's---are exposed as \texttt{CSlice<'a, T>} and \texttt{CSliceMut<'a, T>}, which yield element handles rather than \texttt{\&[T]}/\texttt{\&mut [T]}, so no reference to an element is formed.

\begin{table}
  \caption{\texttt{ffibox} primitives and the standard-library counterparts whose interfaces they mirror.}
  \label{tab:ffibox}
  \small
  \begin{tabular}{@{}lll@{}}
    \toprule
    Challenge & \texttt{ffibox} primitive & Standard counterpart \\
    \midrule
    C1 & \texttt{Foo} (layout newtype) & --- \\
    C2 & \texttt{CBox<Foo, P>} & \texttt{Box<T>} \\
    C2 & \texttt{CArc<Foo, P>} & \texttt{Arc<T>} \\
    C2 & \texttt{CVal<Foo, P>} & \texttt{T} (by value) \\
    C3 & \texttt{FooRef<'a>} / \texttt{FooMut<'a>} & \texttt{\&T} / \texttt{\&mut T} \\
    C4 & \texttt{CGuard} / \texttt{FooLocked<'a>} & \texttt{MutexGuard<T>} \\
    C5 & \texttt{CVec<T, P>} & \texttt{Vec<T>} \\
    C5 & \texttt{CStrBox<P>} & \texttt{CString} \\
    C5 & \texttt{CSlice<'a, T>} / \texttt{CSliceMut<'a, T>} & \texttt{\&[T]} / \texttt{\&mut [T]} \\
    \bottomrule
  \end{tabular}
\end{table}

\paragraph{(C6) Type safety.}
Each obligation is an \texttt{unsafe} trait that a binding implements once per type and that the primitives require as bounds. \texttt{CCell} links a layout type to its C type and to its \texttt{Ref} and \texttt{Mut} handles, so that generic code, such as \texttt{CBox<T, P>} or \texttt{CSlice<'a, T>}, reaches C objects only through handles. The lifecycle traits are implemented on the policy (C2): \texttt{CDrop} releases an object, \texttt{CDupClone} and \texttt{CRefClone} clone it by deep copy or by reference count, \texttt{CDispose} releases the resources of a value held by a \texttt{CVal}, and the construction traits \texttt{CAlloc} and \texttt{CNew} extend \texttt{CDrop}, so every safe path into an owner goes through the policy that releases it. \texttt{CGuarded} registers a type's C lock, and \texttt{CZeroable} asserts that the all-zero bit pattern is a valid value. 

Macros emit the common implementations: \texttt{define\_ctype!} a layout type with its handles, and the policy macros bind a C routine at its exact pointer type, so that a routine for the wrong type fails to compile. Since these macros expand to \texttt{unsafe impl}s, the binding's author need not write \texttt{unsafe} to invoke them, but each invocation carries the trait's proof obligation, e.g., that the bound routine indeed releases or clones the object.

\subsection{ffimap: Mapping Safety on the FFI Surface}
\label{sec:scan}

A binding falls short of its goals at specific sites: public \texttt{unsafe} functions and raw pointers on the safe interface cost ergonomics, and C objects managed outside \texttt{ffibox}'s primitives, e.g., through a hand-written \texttt{Drop}, escape \texttt{ffibox}'s audit. ffimap is a static analyzer that locates these sites. Since its analysis has precision gaps (discussed below), ffimap acts as a linter that sets a target for each property and points to the sites to review, rather than as a verifier.

ffimap is a rustc driver that runs after type checking and reads the crate's \gls{hir} together with its inferred types. It thus counts after macro expansion and \texttt{cfg} selection, resolves type aliases, re-exports, and generic instantiations to the types they denote, and decides visibility by rustc's effective visibility, i.e., whether an item is reachable from outside the crate, rather than by its \texttt{pub} keyword (Appendix~\ref{sec:appendix-ffimap} compares it with syntactic tools). Besides the crate, ffimap takes two lists from its user: the \texttt{-sys} crates, whose \texttt{\#[repr(C)]} types it takes as C's, and the names of the crate's seams, which we define below. Its counters fall into two categories. The \emph{generic} counters apply to any Rust crate, which lets us compare our bindings with community-maintained ones (\S\ref{sec:eval-comparison}). The \emph{\texttt{ffibox}} counters read each type's role from declarations that \texttt{ffibox} and Crustify's conventions add, which we list below. Every count comes with the source sites behind it, and tests and the generated \texttt{-sys} crates are excluded. Agents run ffimap to locate residual \texttt{unsafe} code and raw pointers in their batch, and the assistant records a scan after every batch, wave, and module, from which our evaluation draws its figures (\S\ref{sec:eval}).

\paragraph{Seams.} A \emph{seam} is a function that only converts between a wrapper and the C value it wraps, such as \texttt{from\_raw}, which adopts a C pointer, or \texttt{as\_ptr}, which hands one out. Seams are \texttt{unsafe} or expose a raw pointer by necessity, since no Rust type can express the validity of the pointer they convert, but an ergonomic binding never requires a consumer of its safe API to call one. Seams serve code that itself calls C, i.e., bindings built on other bindings, tests against C, and, as a fallback, consumers that need a C item that the bindings do not cover yet. ffimap counts them apart from the surface that consumers rely on, and takes them by name, since a signature does not show that a function only converts.

\paragraph{Unsafe surface.} The generic counters count the public and private \texttt{unsafe fn}s, trait methods included, the \texttt{unsafe} blocks and the lines they span, the \texttt{unsafe impl}s, and the calls whose callee carries a foreign ABI, e.g., \texttt{extern "C"}, including calls through function pointers. They split the public \texttt{unsafe fn}s by what made them \texttt{unsafe}: a method of a trait implementation whose trait declares the method \texttt{unsafe}, such as \texttt{CDrop::c\_drop}, which an implementation cannot declare safe; a seam; and the remainder, which the crate chose to make \texttt{unsafe}.

\paragraph{Raw pointers.} The generic counters count every \texttt{*const T} and \texttt{*mut T} in a signature or a field, including \texttt{NonNull<T>} and \texttt{AtomicPtr<T>}, which wrap one, and pointers nested in slices and generic arguments. Each position is classified by the visibility and safety of its function, or by the visibility of its type and field, so that the pointers on the \emph{safe public interface}, i.e., in the public fields of exported types and in the signatures of exported safe functions, are told apart from those that only \texttt{unsafe} callers or the crate itself reach. The positions in the methods of trait implementations are counted apart, since the trait rather than the crate fixed them, and so are those of seams.

\paragraph{\texttt{ffibox} declarations.} Without \texttt{ffibox}, a newtype that covers a C struct, a borrowed handle to one, and a Rust value type look alike, and a function that wraps a C function cannot be told from a helper. The \texttt{ffibox} counters resolve these cases from four declarations. (i)~A \emph{layout type} is a type that implements \texttt{CCell} (C6), whose associated types name the C struct it covers (\texttt{CCell::C}) and its shared and exclusive handles (\texttt{CCell::Ref}, \texttt{CCell::Mut}). (ii)~A \emph{policy} is a type that implements one of \texttt{ffibox}'s lifecycle traits (C6). (iii)~An \emph{owner} or a \emph{borrow} is one of \texttt{ffibox}'s owning or borrowing primitives (Table~\ref{tab:ffibox}). (iv)~A \emph{wrapper} of a C function or field is an item that carries a rustdoc alias naming it, e.g., \texttt{\#[doc(alias = "git\_repository\_open")]} or \texttt{\#[doc(alias = "git\_oid.id")]}, which Crustify's conventions place on every item that stands in for a C item, and which ffimap resolves against the \texttt{-sys} crates' declarations.

\paragraph{\texttt{ffibox} use.} From these declarations, the \texttt{ffibox} counters inventory the C surface: the functions that the \texttt{-sys} crates declare, those that the bindings call or take the address of, and those that a wrapper names; and the structs that a \texttt{-sys} declaration puts behind a raw pointer, those that the bindings name, and those that some \texttt{CCell::C} names. For each type, they record which of \texttt{ffibox}'s traits it implements and whether a macro of \texttt{ffibox}, such as \texttt{define\_ctype!}, emitted each implementation, which its source span shows. In public signatures and fields, they count the owners and borrows taken and returned, and classify each mention of a layout type \texttt{Foo} by its innermost holder: an owner (\texttt{CBox<Foo, P>}), a borrow (\texttt{CSlice<'a, Foo>}), a standard owner (\texttt{Box<Foo>}, \texttt{Vec<Foo>}), a reference (\texttt{\&Foo}, \texttt{\&mut Foo}, \texttt{\&[Foo]}), or none, i.e., \texttt{Foo} by value. Mentions behind a raw pointer or in \texttt{PhantomData} are not counted.

\paragraph{Outside \texttt{ffibox}.} A binding bypasses \texttt{ffibox} wherever it manages a C object or a pointer by hand, or forms a reference that \texttt{ffibox}'s invariant rules out. The generic counters find the types that implement \texttt{Drop} by hand, or a reference-forming trait, i.e., \texttt{Deref}, \texttt{AsRef}, \texttt{Borrow}, \texttt{Index}, or their mutable counterparts, and the types that store a raw pointer, both those that declare one in their own fields and those that reach one through their crate's or the \texttt{-sys} crates' types, but not inside another crate's owner, such as \texttt{Vec} or \texttt{CBox}. The \texttt{ffibox} counters single out the raw-pointer types that also implement one of \texttt{ffibox}'s traits, and thus declare the pointer's lifecycle through it. They also count the \texttt{-sys} structs that the bindings use without a layout type or hold raw, by value, in public signatures, and the references \texttt{\&Foo}, \texttt{\&mut Foo}, and \texttt{\&[Foo]} to a layout type \texttt{Foo} in public signatures and fields, except for the receivers that \texttt{ffibox} sanctions: those of \texttt{Foo}'s inherent \texttt{as\_ref} and \texttt{as\_mut}, which turn a \texttt{Foo} that Rust holds inline (C1) into its handles (C3), and those of trait methods, whose shape the trait fixes, e.g., \texttt{Clone::clone(\&self)}. Only the receiver is exempt, so a trait method's other arguments and its return count like any function's.

\paragraph{Known limitations.} ffimap is neither sound nor complete. It misses the references that an autoref through a raw pointer forms, e.g., \texttt{(*p).field}, since neither a signature nor a cast carries them, and the references to \texttt{-sys} structs that no layout type covers, since it finds a struct's mentions through \texttt{CCell::C}. Conversely, it counts every reference to a layout type outside the sanctioned receivers, even where Rust holds the value, e.g., \texttt{\&Oid}. Its \texttt{ffibox} counters are only as accurate as the declarations they read: seams, \texttt{as\_ref}, and \texttt{as\_mut} are matched by name, so that a conversion named otherwise is counted like any function, and a wrapper without its rustdoc alias is missed.

\subsection{Wavefront: Dependency-Ordered Scheduling}
\label{sec:wavefront}

Wrapping a repository-level target requires ordering its types and symbols such that every item is wrapped after the items it depends on. In C, this order often forms cycles and is partly hidden from type declarations, in casts that implement polymorphism, type-erased \texttt{void *} slots, and macros that implement generic types (\S\ref{sec:intro}). Wavefront therefore combines a symbolic analysis with \gls{llm} judgment: its graph is a deterministic function of the entities (T1) and relations (T2) that it extracts from the code, and of a small table of corrections (T3) that agents author for the relations that require judgment. We describe Wavefront along four challenges (D1--D4).

\paragraph{(D1) Dependency graph.}
Wavefront extracts entities and relations with 25 queries over a CodeQL~\cite{codeql} database of the target, i.e., a relational representation of the program that CodeQL records from its build and that Datalog-based QL queries~\cite{ql2016} analyze, and composes them into a single directed graph whose nodes are types (structs, unions, and enums) and symbols (functions, globals, macros, and callbacks). An edge $A \to B$ then states that $A$ requires $B$ to be emitted first. Type-to-type edges follow struct field types; symbol edges follow the types in a symbol's signature and body, its callees, and the globals it accesses. Function-pointer typedefs are treated as symbols, reached through the signatures and indirect calls that use them. Typedef chains are resolved to their underlying tags, so that a type and its aliases form one node. Lifecycle routines (constructors, destructors, and cloners) remain ordinary symbol nodes rather than being attached to their types: C does not distinguish them from a type's other functions or from free functions, so only agents can identify them, and the graph does not depend on their findings.

\paragraph{(D2) Type erasure and polymorphism.}
Wavefront recovers the relations that casts and type-erased slots hide from the conversions that the code writes, without data-flow analysis. A function whose body reads its \texttt{void *} parameter as a type \texttt{T} depends on \texttt{T}, as on a typed parameter. A type stored into or read from a type-erased field is a \emph{variant} of the field's owner, and depends on the owner as a derived type on its base, so that the owner can become generic over its variants. A cast between two named types, written directly or through one slot, is an edge in its written direction, so that an upcast orders the base first; where cast edges form a cycle, an edge survives only toward the type that is cast to or from strictly more types, so that the edges run from the instances of a family up to its hub. Each conversion is recorded at the access that writes it, and nothing follows a value through locals, memory, or calls. This keeps the extraction cheap and predictable, but misses relations, e.g., a \texttt{void *} handed on through a chain of calls, and adds spurious ones, e.g., a slot that one file reuses for an unrelated type.

The relations that require judgment are left to agents: the planner follows each unresolved type-erased value hop by hop and records its conclusions as T3 rows that add a missed relation, drop a derived cast edge, fix the direction of a base--derived pair, or name a macro as the generator of a family of types, each with a report that gives its evidence. Wavefront checks each row only for well-formedness against the extracted facts, e.g., that the entities it names exist, and not the relation itself. Rows can, however, only add or remove cast and erasure edges, never the field, signature, and body edges that follow from declarations, so that a wrong row can misorder a family of related types, but cannot schedule an item before a type that its fields or signature need. In the public-signature graph, cast edges are added only between items that fields and signatures already reach, so that a cast to an internal type never pulls that type into the schedule. The graph thus orders each family of related types after its root, i.e., its base, the owner of the field its members are stored in, or the macro that generates them, so that Crustify can bind the root first as a generic Rust type or trait (\S\ref{sec:synthesis}).

\paragraph{(D3) Ordering and batching.}
Wavefront orders the target's \emph{units}, i.e., its types and symbols, except enums and macros other than type generators. It proceeds in five steps. (1)~\emph{Grouping.} Wavefront groups the units by the modules that its configuration lists and condenses the graph into a module graph. An edge between two modules weighs the distinct units that one uses of the other. Without modules, each unit forms its own group, and the order is flat. (2)~\emph{Cutting cycles.} Wavefront finds the \glspl{scc} of the module graph~\cite{tarjan1972} and orders each one's members by how many more of its edges reach them than leave them. It counts each dependency once, since cutting it leaves one binding to write. Every edge to a later member becomes a cut. The order is deterministic, but not a minimum cut. (3)~\emph{Layering.} Wavefront assigns each module to a layer by its longest dependency path, as in Sugiyama et al.~\cite{sugiyama1981}. It then repeats steps 2 and 3 on the units of each module. (4)~\emph{Recording cuts.} Each cut is recorded at both of its ends, and a cut between modules also lists the unit edges it crosses. Within a module, Crustify binds a cut dependency to its raw \gls{ffi} type, and the batch that later wraps it upgrades the binding. Between modules, Crustify wraps the crossing units before their consumer (\S\ref{sec:planning}). (5)~\emph{Waves and batches.} Each layer of a module's units forms a \emph{wave}, which records the waves it depends on. Crustify runs each module's waves one at a time, modules that do not depend on each other concurrently, and the batches of a wave concurrently (\S\ref{sec:synthesis}). Type batches are bounded by a number of types and close early once their declared fields reach a threshold. Symbol batches are bounded by a number of symbols and lines of function bodies. Adjacent layers of one module whose units fit into one batch are merged. Since Wavefront always orders the whole target, a schedule that covers only part of it is part of the same order, so separate planning sessions produce consistent schedules.

\paragraph{(D4) Scoping the workload.}
Wavefront takes as input a JSON configuration with three parts. The \emph{closure} names the program that it analyzes: the target's implementation files, i.e., its sources and private headers, and the API headers that publish its interface, each with paths to exclude. The optional \emph{modules} group the closure's units for ordering (D3). The optional \emph{seeds} select the files, or the types and symbols within them, that a schedule starts from. Wavefront builds its graph either over the whole implementation or over the published API only: for wrapping, it traverses a public-signature graph that omits function bodies, retains the fields of public definitions, and keeps forward-declared types opaque. Items that the closure uses but does not define, e.g., from libc, are \emph{external}: they remain dependencies of the units that reach them, but are never units themselves, and Crustify wraps them in campaigns of their own (\S\ref{sec:planning}). A schedule holds its seeds and every unit they depend on, so that the schedule of one module, or of a few named types and symbols, is complete on its own.

\paragraph{Queries.}
Wavefront answers agents' queries from the same relations it schedules from: a type's fields and the functions that access each field, a function's callers and callees, including calls through callbacks, the symbols that take a given type, the functions that take or return a type-erased or string pointer, optionally restricted to those that reach a given routine within a number of calls, which locates candidate lifecycle routines, and the lifecycle routines that agents have identified for a type. Coding agents otherwise search the code as text~\cite{claudecode}, which misses uses that a macro expansion or a typedef alias hides and conflates same-named fields of different structs.

\paragraph{Accuracy.}
Wavefront's guarantees are relative to its inputs: CodeQL resolves the graph from the compiler's view of the build, so code that the build configuration excludes is not seen. The schedule is a deterministic function of the extracted facts, the T3 rows, and the configuration; its seeds select part of the order but never change it, and ownership findings never move a node. Its only approximate part is the cast and erasure edges (D2), whose errors may keep an agent from recognizing a family and binding it generically, but cannot invalidate the order of the members' fields and signatures.

\paragraph{Ownership store.}
Wavefront records the ownership and lifecycle properties that agents establish in an \emph{ownership store}. Each pointer in a field, an argument, a return value, or a global is recorded with the properties that select its \texttt{ffibox} representation (\S\ref{sec:ffibox}): its cardinality (a single object, an array, or a string), whether it is owned or borrowed and from which source, whether it may be null, and whether its pointee is written through it. A pointer that is both owned and borrowed has \emph{dual ownership}, i.e., it owns its pointee on some paths and borrows it on others; for an out-parameter or an array of pointers, the record also gives the ownership of the inner pointee. A symbol records its \emph{lifecycle role} for one of its pointers: a \emph{dropper} releases an object and its storage, a \emph{disposer} only the resources it holds, a \emph{cloner} copies it deeply or by reference count, and a \emph{constructor} allocates or initializes it. A type may thus have several routines of each kind. Fields may further be marked as reference counts or as protected by a lock, and callbacks may be forked by call site. Wavefront checks each submission for consistency with the record that it derives from the code and with the rules of each role, e.g., that a destructor's argument is owned and a cloner's borrowed, and rejects the whole submission otherwise; like the checks of T3 rows, these rule out malformed findings, not wrong ones. Appendix~\ref{sec:appendix-examples} shows how such findings map to \texttt{ffibox} primitives.

\subsection{Workflow}
\label{sec:crustify-workflow}

There is no established methodology for generating safe bindings for an entire repository. Crustify therefore defines one in a \emph{specification}, a set of descriptive documents that state what each artifact of a campaign holds, how each decision is made, and what each procedure produces, without naming who executes a step. It comprises workflows for whole campaigns, how-tos for single items, and schemas for the artifacts that steps exchange, and states each rule once, so that a change reaches every agent; role prompts assign its steps to agents (\S\ref{sec:agents}). A wrap campaign proceeds in four phases: \emph{setup} establishes the campaign's scope and prepares the target; \emph{planning} decomposes it into modules and schedules the workload in dependency order; \emph{binding} generates the bindings, writes coverage-guided tests to reveal their flaws, and runs the regression suite after every batch; and \emph{review} audits the bindings of each wave before they are promoted. Since the workload is scheduled per batch, a target can be wrapped incrementally, so that maintainers can ship bindings at intermediate milestones rather than only once the entire repository is wrapped.

\subsubsection{Setup}
\label{sec:setup}

The campaign's scope, i.e., the entire public API, some of its modules, or named types and symbols, is settled with the user before setup, in the campaign's settings and brief; setup then prepares the target for the phases that follow.

\paragraph{Instrumented builds.}
The target library is built once in several configurations, which all later steps reuse: a plain build, which serves as the functional baseline; a coverage-instrumented build, which measures C coverage during testing; an \gls{asan}/\gls{ubsan} build for memory-safety and \gls{ffi} tests; a separate \gls{tsan} build for race tests, since \gls{tsan}'s runtime conflicts with \gls{asan}'s; and a BorrowSanitizer build for aliasing violations. Our campaigns omit \gls{msan}, which reports false positives unless every part of the program is instrumented, including prebuilt shared libraries, and Miri needs no C build, as it interprets only the Rust code. The library's own test suite is run on the plain build to record a baseline that the final bindings must preserve; tests that are unstable or need infrastructure the host lacks, e.g., a network service, are each disabled with its reason. Finally, a CodeQL database is created from a traced plain build, from which Wavefront extracts its entity and relation tables (\S\ref{sec:wavefront}).

\subsubsection{Planning}
\label{sec:planning}

Planning decomposes the target into modules and schedules their types and symbols in dependency order. The decomposition follows the codebase's own organization rather than a judgment of how it should be organized, so that the Rust tree mirrors it.

\paragraph{Module decomposition.}
The repository is decomposed into the link units that its build produces, i.e., shared objects, static libraries, and executables, as its linker outputs show, and each link unit into \emph{modules}: a module is a single translation unit together with the headers that carry its name, internal or installed. A shared header, which carries no single translation unit's name, is split item by item, each item going to the translation unit that implements its lifecycle or logic. Every translation unit that the build compiles in scope thus belongs to exactly one module. This step produces \texttt{modules.json}, which records the link units, their modules, and the dependencies between modules, including the kinds of items consumed across them. The out-of-tree libraries that the target links, e.g., libz, are recorded once in \texttt{dependencies.json}, with the release that the target links and its source, and each module records what it consumes from them. Decomposition also identifies the \emph{foundational} modules, i.e., the translation units that define the allocators, releasers, disposers, and cloners of type-erased memory, and those whose string routines build only on these. Every other module may depend on them, and their routines are bound as \texttt{ffibox} policies through which the other modules' bindings allocate and release their objects.

\paragraph{Scheduling.}
Modules are ordered bottom-up by their dependencies. The foundational modules depend only on each other and run first, those of type-erased memory before those of strings, which call them. Wavefront orders the modules and cuts their cycles (D3, \S\ref{sec:wavefront}); the units that a cut dependency crosses are wrapped before their consumer rather than reached through raw bindings. The out-of-tree libraries in \texttt{dependencies.json} are wrapped lazily: before a wave that uses entities a library's wrapper lacks, the wrapper's own campaign binds just those, and the target's workspace pins the result, so that a wrapper covers only what its consumers use and later campaigns extend it. Within each module, the types and symbols are scheduled into waves of size-bounded batches of mutually independent items (D3--D4), so that type batches land before the symbol batches that depend on them. The order also follows the casts and type-erased slots of D2, so that the root of each family of related types is bound first. The schedule is recorded in \texttt{schedule.json}.

\paragraph{Rust tree.}
Once the plan is complete, the Rust tree is scaffolded for the link units in scope: the safe crate, with one Rust module per link unit, and one \texttt{-sys} crate per link unit, whose bindgen~\cite{bindgen} input covers the target's C code and whose allowlists the batches fill as they bind items. Everything below a link unit, e.g., a module's source file, is created lazily, before the module's first wave.

\subsubsection{Binding}
\label{sec:synthesis}

The binding phase runs each module's waves one at a time, in schedule order, and the batches of each wave concurrently. Modules that do not depend on each other run concurrently, each on a branch of its own that is integrated into the working branch once its last wave has been promoted, so that each module finishes, and can ship, as early as its dependencies allow. Each batch wraps its items with \texttt{ffibox} primitives, refines them through compiler and lint feedback and feedback on their \texttt{unsafe} surface (\S\ref{sec:scan}), and tests them (\S\ref{sec:testing}). The central decision in every batch, whether it wraps types or symbols, is how to represent each pointer that is exposed on the safe API or crosses the \gls{ffi} boundary.

\paragraph{Pointer and lifecycle analysis.}
Each batch starts by establishing the ownership and lifecycle properties of its items, i.e., the properties that select each pointer's representation and the routines that construct, release, and clone each type, and records them in the ownership store (\S\ref{sec:wavefront}) for later batches and review. C declares few of these properties, so the batch derives them from every path that stores, transfers, clones, or frees a pointer, rather than from names and comments, which it overrides where the code disagrees. The findings guide the batch's representations, accessors, and tests, but no step generates code from them.

\paragraph{Pointer representation.}
Crustify aims to represent each \gls{ffi} pointer with an \texttt{ffibox} primitive (\S\ref{sec:ffibox}), so that the compiler checks its ownership, mutability, nullability, and cardinality. C programs, however, settle these properties by convention, and the same pointer often carries different semantics depending on how a function uses it or on a flag that the caller passes; ffmpeg's \texttt{av\_dict\_set}, for instance, takes ownership of the strings it receives only if the caller sets a flag. A Rust signature that mirrors the C one would either be unsound or shift the distinction onto the caller as an \texttt{unsafe} contract. Crustify instead \emph{forks} such pointers: it emits one Rust form per semantics that the analysis reveals, each of which fixes ownership, mutability, nullability, and cardinality in its type, e.g., an owned and a borrowed form for a pointer with dual ownership, so that the compiler checks every use. Forking applies to struct fields, function signatures, and callbacks alike. For type-erased pointers, Crustify additionally makes owners and borrowed handles generic over the pointee, bound by \texttt{CCell}, so that only types with an \texttt{ffibox} layout, which never forms a reference into C memory, can stand in for it. Appendix~\ref{sec:appendix-examples} gives examples of the bindings that the binding phase emits, mapped to the \texttt{ffibox} primitives they use.

\paragraph{Types.}
A type batch contains structs and \emph{type-generating macros}, i.e., macros whose expansion defines an entire aggregate type, and wraps each type's layout and field accessors together with its lifecycle routines, i.e., its constructors, destructors, and cloners. A union is bound in the batch of the struct that holds its discriminator, as a Rust enum tagged by it when the discriminator determines the variant, and enums are imported from the raw bindings as constants. The batch emits a layout newtype over the type's bindgen representation, together with its borrowed handles, usually through \texttt{ffibox}'s \texttt{define\_ctype!} macro. Layouts and handles that carry lifetimes, e.g., a handle that borrows both an object and the object it was looked up in, are written by hand, since the macro covers only a single handle lifetime. The batch also emits accessors for the fields that the schedule lists, i.e., every field of a type that the API headers define, and otherwise those that its consumers access: a getter and a setter for each, forking pointer fields as described above: an owned field gets a setter that releases the previous owner and getters that either take or borrow it, and a borrowed field ties its value to the layout's lifetime. Families of related types are bound around their root, which the schedule orders first (\S\ref{sec:planning}). A polymorphic base becomes a generic type bounded by a marker trait, and each derived type implements the trait, so that a derived object is used where its base is expected without unsafe casts or transmutes. A type-erased field likewise becomes generic, bounded by \texttt{CCell} and, when the types stored in it are known, by a marker trait that these types implement. A type-generating macro becomes a generic Rust abstraction that its instances implement, bound by hand, since \texttt{define\_ctype!} covers a single C type.

\paragraph{Foundational modules.}
The batches of the foundational modules (\S\ref{sec:planning}) run before those of any other module, since their routines allocate, release, and copy the plain \texttt{void *} and \texttt{char *} buffers and strings that the other modules use. Each routine with a lifecycle role becomes an \texttt{ffibox} policy that calls it through the \gls{ffi}, rather than an ordinary safe function: the allocator, releaser, and cloner that manage the same memory, e.g., \texttt{av\_malloc}, \texttt{av\_free}, and \texttt{av\_memdup}, form one policy, which backs \texttt{CBox} owners of single objects and \texttt{CVec} owners of arrays, or one policy each, and the policy of the string routines backs \texttt{CStrBox} (Appendix~\ref{sec:appendix-examples}). Every later batch reuses these policies, so that the buffers and strings it binds are released through the same routines as in C. Allocation is otherwise enclosed in each type's constructor, bound through \texttt{ffibox}'s \texttt{CNew} trait, and generic allocators are exported only for objects whose type no C constructor allocates; these return an initialized object only for types that implement \texttt{CZeroable}, and \texttt{MaybeUninit<T>} storage otherwise.

\paragraph{Symbols.}
A symbol batch contains functions, globals, and callbacks, i.e., function-pointer typedefs. Macros are not published as Rust APIs of their own: an alias is replaced by its symbol's wrapper, a function-like macro is bound through the smallest C shim it needs, and a constant macro uses its bindgen declaration. For each function, the batch emits a wrapper whose arguments and return value are safe handles; the wrapper reconstructs raw pointers only at the \gls{ffi} call, inside a small \texttt{unsafe} block that documents why the call is sound. A function whose pointer arguments or return value admit several semantics is forked into one wrapper per semantics, as described above, and a function that takes or returns a type-erased value becomes generic over the types it is read as, rather than exposing a \texttt{void} pointer. A function that operates on a type becomes a method of \texttt{FooRef}, \texttt{FooMut}, or \texttt{FooLocked}, as it reads, mutates, or requires the lock of the object, and returns values bound to the object's lifetime rather than to the handle's borrow, which lifetime elision would choose. A wrapper is declared \texttt{unsafe} only when no type can express the caller's obligation.

A global is wrapped as a \texttt{'static} borrow, never as an owner, since its storage is never released. Its access discipline follows from all of its accesses in C: a global that is immutable after initialization gets a plain shared reference, the exception of \S\ref{sec:ffibox}; one that C protects with a lock is reached through \texttt{ffibox}'s \texttt{CGuardedRef}, which hands out a handle only while it holds the lock; a thread-local global gets a handle that is borrowed only on the current thread; and any other mutable global gets \texttt{unsafe} accessors whose contract excludes concurrent access.

A callback becomes a newtype over the raw function pointer, whose \texttt{call} method converts safe arguments to raw ones and the result back. When the functions that invoke a callback type realize different ownership contracts for its arguments, e.g., some transfer an argument's ownership to the callback while others lend it, the callback is forked into one wrapper per contract, which the ownership store records with the invokers that realize it.

\paragraph{Testing.}
\label{sec:testing}
Each batch writes four kinds of tests for the items it wraps. \emph{Equivalence} tests run the C implementation and the Rust API on independent but equivalent inputs and compare return values, out-parameters, buffers, callback invocations, and lifecycle effects after each call of a multi-call sequence. \emph{Safe \gls{ub}} tests are compiled with \texttt{\#![forbid(unsafe\_code)]} and drive the safe API, including with programs that the compiler must reject, and must not trigger any sanitizer; \emph{unsafe \gls{ub}} tests call each public \texttt{unsafe} function within its documented contract; and \emph{unit} tests check internal routines. Each prepared instrument is a separate obligation, e.g., use-after-free and double frees under \gls{asan}, races through asserted \texttt{Send} and \texttt{Sync} implementations under \gls{tsan}, \gls{ub} within the Rust code, including aliasing violations, under Miri, and aliasing violations across the boundary under BorrowSanitizer. The batch then measures the line coverage of its items in C, with the coverage-instrumented build, and in Rust, adds tests for reachable uncovered paths, and reports the paths it deems unreachable. Coverage counts only non-test code, since tests would otherwise be credited for executing themselves; each kind of test is measured alone, and the total is the union of their covered lines.

Every batch, whether it wraps or reviews, ends with a regression gate: it rebuilds the workspace, runs the linter, and runs the full test suite under \gls{asan}, \gls{ubsan}, and \gls{tsan}, against the matching sanitized C build. Two instruments run less often: Miri runs once per wave, over the whole workspace, in the wave's review; and BorrowSanitizer, the slowest instrument, which in our experience also reports false positives at the foreign boundary, runs only on the tests that a batch expects to trigger it, and over the whole workspace in the review of a module's last wave (\S\ref{sec:review}).

\subsubsection{Review}
\label{sec:review}

Each binding wave is followed by a review wave, which audits the landed batches for undefined behavior, equivalence drifts, and coverage gaps before any dependent wave starts. We believe that review is crucial between waves, because a flawed binding propagates into the bindings that depend on it, so that fixing it later requires larger, breaking changes to their APIs. Each review batch covers the commits of one or more binding batches of the reviewed wave, which it reviews for the objectives that the brief selects, against the same binding how-to that the binders follow; its reviewer can run on a different model than the binders whose work it audits (\S\ref{sec:eval-setup}). Each review wave also runs Miri over the whole workspace, and the review of a module's last wave runs BorrowSanitizer over everything that the module landed, once on the wave as landed and once on the reviewed tree. Each defect found is filed as a report, and fixed under the \emph{fix} objective, which our campaigns select, in one of eight categories: \emph{lifecycle}, a missing lifecycle routine or owner; \emph{unsafe}, an \texttt{unsafe} function or raw pointer on the public API that a safe binding could replace, located with ffimap; \emph{obligations}, a safety obligation or its justification that is wrong or incomplete; \emph{UB}, through the safe or the \texttt{unsafe} API; \emph{equivalence}, a divergence from the C implementation; \emph{internal}, a defect in private code that the unit tests missed; \emph{conventions}, a deviation from Crustify's coding conventions; and \emph{misc}, any other defect. A \gls{ub} or equivalence candidate becomes a finding only if it reproduces through the Rust API, as a standalone reproducer that triggers a sanitizer or fails an equivalence comparison; the reproducer is committed before its fix, so that it can be rerun against the unfixed code, and the fix turns it into a regression test. Coverage gaps are filled with tests rather than reports, since the coverage measurements show the difference. Reviewers also verify the ownership and lifecycle findings of the reviewed items, e.g., that no constructor, destructor, or cloner of a type was missed, and resubmit each finding they correct or add to Wavefront, with a report of the evidence. A reviewed wave is promoted once its review batches have landed.

\subsection{Multi-Agent Architecture}
\label{sec:agents}

Crustify executes the workflow with four agent roles: an \emph{assistant}, Crustify's co-translator, which interacts with the user and owns the campaign, and \emph{planner}, \emph{binder}, and \emph{reviewer} agents, which it spawns. The assistant executes setup (\S\ref{sec:setup}), planner runs execute planning (\S\ref{sec:planning}), binders execute the binding phase (\S\ref{sec:synthesis}) and reviewers review (\S\ref{sec:review}), one batch each, and the assistant owns every step in between, e.g., scaffolding the Rust tree and preparing, landing, and promoting each wave. Each agent runs as a session of an off-the-shelf coding agent, e.g., Claude Code or Codex, and roles can use different models. Only the assistant runs interactively; every other agent runs headless in its own git worktree and branch, and sees only its own worklist. A binder or reviewer lands its commits on its wave's branch one batch at a time, under a lock: it rebases onto the branch, reruns the gates that the rebase could affect, and fast-forwards the branch, so that concurrent batches never overwrite each other's work.

\paragraph{Prompts.}
Each agent's prompt consists of three layers: the \emph{specification} (\S\ref{sec:crustify-workflow}), included whole in every role that follows it; the \emph{role prompt}, which states the role's inputs, the steps it owns, and the commands and judgments that the specification leaves to it; and \emph{tool specs}, which add one tool each, i.e., \texttt{ffibox}, Wavefront, or ffimap: a header that every role includes, and one file per specification document that the tool serves, which follows that document wherever it is included, so that removing a tool's specs removes it from every role while the specification stays unchanged. The prompts state every step, so that an agent neither omits one nor drifts from it over a long run, but contain no code examples and leave the execution to the agent. Crustify places the whole prompt in the coding agent's system prompt, which the agent does not compact when its context fills up (\S\ref{sec:bg-agents}). When a task needs more than its role's prompt covers, e.g., after the user changes the campaign's direction, the assistant writes \emph{additional instructions} for that agent, which take precedence over the rest of its prompt and name every departure from the specification explicitly, so that the agent does not reconcile it away.

\paragraph{Assistant.}
The assistant is the user's entry point. Started with an instruction to run the campaign, it executes the workflow end to end; otherwise, it serves the user's requests, using the workflow as a reference, e.g., to plan a single module and stop. Its sessions are resumable, and it rereads the settings before each module, so that the user can change course while the campaign runs. Between approval gates, it runs autonomously: it runs and promotes the planner runs, scaffolds the Rust tree, prepares each wave, spawns one binder or reviewer per batch up to the approved parallelism, promotes completed waves, and keeps the campaign's accounting, but binds and reviews no batch itself, and advances the shipping branch only with the user's approval. When a campaign exposes a defect in Crustify or one of its tools, the assistant fixes it on a separate branch of that component, and before each wave it curates a shared agent memory, through which agents share lessons that need no code change.

\paragraph{Planner.}
Planner runs execute planning one after another, each on a branch of its own. Each run brings the committed plan up to date and corrects what earlier runs decided, including Wavefront's correction tables (\S\ref{sec:wavefront}), filing a report with the evidence for each correction. The assistant checks each run's changes before promoting it, e.g., that no frozen module or started wave changed, and re-plans a changed scope or workload the same way, without re-planning landed work.

\section{Evaluation}
\label{sec:eval}

We evaluate Crustify on two campaigns that wrap the public APIs of libgit2 and ffmpeg, and answer the following research questions:
\begin{description}
  \item[RQ1 (Safety and ergonomics).] How much \texttt{unsafe} code and how many raw pointers remain on the bindings' public API, and how consistently do the bindings build on \texttt{ffibox}'s primitives rather than their own, compared with community-maintained bindings?
  \item[RQ2 (Soundness evidence).] How thoroughly do the bindings' tests exercise Rust and C code under sanitizers and Miri, and what defects does review find, compared with community-maintained bindings?
  \item[RQ3 (Coverage and cost).] How much of a repository-level public API does Crustify wrap, at what cost in money, tokens, and time?
  \item[RQ4 (Dependency-ordered scheduling).] What schedules does Wavefront produce, and how many of the dependency relations that casts and type-erased slots hide do agents correct?
  \item[RQ5 (Ownership and lifecycle reasoning).] Which ownership and lifecycle contracts do the agents derive for the pointers and types that they wrap, and how accurate are they?
\end{description}

\subsection{Setup}
\label{sec:eval-setup}

Both campaigns target every non-deprecated item exported by their target's public headers, i.e., libgit2's and those of ffmpeg's seven libraries. We count API coverage over these items except enums, whose constants the bindings import from the raw bindings (\S\ref{sec:synthesis}), over \lgApiTypes{} types and \lgApiSyms{} symbols for libgit2, and \ffApiTypes{} and \ffApiSyms{} for ffmpeg. They run with at most 2 types per type batch, which closes once its types' declared fields reach 20, so that a type with 20 or more fields is scheduled in a batch of its own, and with at most 25 symbols and 500 lines of function bodies per symbol batch, where a longer symbol is scheduled in a batch of its own. Up to 16 batches run concurrently, and up to 32 in waves of small review batches, on a machine with an Intel Xeon CPU (144 hardware threads) and 1\,TB of RAM. libgit2's C code is built with Ubuntu clang 18.1.3 and ffmpeg's with gcc 14.2 in a plain configuration, and with Ubuntu clang 18.1.3 and 20.1.2 respectively with \gls{asan} and \gls{ubsan}, and with \gls{tsan}, while the BorrowSanitizer configuration uses BorrowSanitizer's own clang toolchain and compiler plugin. The coverage we report comes from a separate build of each C library, instrumented with LLVM's source-based coverage (\texttt{-fprofile-instr-generate -fcoverage-mapping}) by clang~23, the LLVM release of the Rust nightly that instruments the bindings, so that one \texttt{llvm-cov} reads both languages. 

One developer supervised both campaigns, which ran concurrently, through the assistant, while most of each campaign ran autonomously. Each campaign ran its modules in \emph{sub-campaigns}, i.e., groups of related modules, such as libgit2's \texttt{index} or ffmpeg's \texttt{crypto} modules, 9 in libgit2 and 8 in ffmpeg, including the foundational ones. The developer approved each campaign's brief and checked the bindings between sub-campaigns, but did not intervene in the agents' work. They occasionally corrected defects in Crustify's prompts and tools, which then did not recur.

Binders and reviewers mainly ran Claude Opus~5.5. GPT-6.1 Sol wrapped libgit2's \texttt{config} and ffmpeg's \texttt{crypto}, GPT-6 Astra wrapped most of libgit2's first \texttt{index} wave and reviewed one ffmpeg batch, and Claude Fable~5.1 reviewed the early sub-campaigns; we stopped using the GPT models since their safety guardrails kept interrupting agents that write tests and reproducers for memory-safety violations. In 64 of the 106 review batches, the reviewer ran on a different model than the binders of its wave. Crustify's prompts and gates were still being corrected while the campaigns ran. We coarsened review from one review batch per binding batch, most of which reported no \gls{ub}, to one review wave per binding wave (\S\ref{sec:review}), moved the regression gates from the assistant to the binders and reviewers, and added the workspace-wide Miri and BorrowSanitizer gates of \S\ref{sec:review}.

\paragraph{Baselines.}\label{sec:eval-comparison} RQ1--RQ3 each compare the generated bindings with the community-maintained git2-rs~\cite{git2rs} for libgit2, and ffmpeg-the-third~\cite{ffmpegthethird} and ffmpeg-next~\cite{ffmpegnext} for ffmpeg, which their maintainers developed over several years. We measured each crate at its latest revision, against the C release it targets (libgit2 v1.9.6 and ffmpeg 9.0), measured coverage as in the campaigns, over the same C, and ran ffimap on each crate, with the seams that its own conventions name (Table~\ref{tab:baselines}).

\begin{table}
  \caption{Generated versus community-maintained bindings. Baseline API coverage is an upper bound that counts every C item the crate's source names, enums excluded as for ours; ours counts wrapped items only. libgit2's C line coverage is over libgit2 with its bundled xdiff and llhttp, which both crates compile, against \lgCovC\% over libgit2 alone (Table~\ref{tab:coverage}). Public functions include the methods of trait implementations.}
  \label{tab:baselines}
  \small
  \setlength{\tabcolsep}{4pt}
  \begin{tabular}{@{}lrrrrr@{}}
    \toprule
    & \multicolumn{2}{c}{libgit2} & \multicolumn{3}{c}{ffmpeg} \\
    \cmidrule(lr){2-3}\cmidrule(l){4-6}
    & Crustify & git2-rs & Crustify & ffmpeg-the-third & ffmpeg-next \\
    \midrule
    API coverage (\%) & \lgApiCov & $\le$\gitApiCov & \ffApiCov & $\le$\fttApiCov & $\le$\nextApiCov \\
    Tests & \lgTests & \gitTests & \ffTests & \fttTests & \nextTests \\
    Rust line coverage (\%) & \lgCovRust & \gitCovRust & \ffCovRust & \fttCovRust & \nextCovRust \\
    C line coverage (\%) & \lgCovCCmp & \gitCovC & \ffCovC & \fttCovC & \nextCovC \\
    Public functions & \lgPubFns & \gitPubFns & \ffPubFns & \fttPubFns & \nextPubFns \\
    \quad \texttt{unsafe} by choice & \lgUnsafeFnOther{} (\lgUnsafeFnOtherPct\%) & \gitUnsafeFnOther{} (\gitUnsafeFnOtherPct\%) & \ffUnsafeFnOther{} (\ffUnsafeFnOtherPct\%) & \fttUnsafeFnOther{} (\fttUnsafeFnOtherPct\%) & \nextUnsafeFnOther{} (\nextUnsafeFnOtherPct\%) \\
    \quad with a raw pointer & \lgRawPtrFns{} (\lgRawPtrFnsPct\%) & \gitRawPtrFns{} (\gitRawPtrFnsPct\%) & \ffRawPtrFns{} (\ffRawPtrFnsPct\%) & \fttRawPtrFns{} (\fttRawPtrFnsPct\%) & \nextRawPtrFns{} (\nextRawPtrFnsPct\%) \\
    \qquad safe, taking one & \lgRawPtrFnsSafeTake & \gitRawPtrFnsSafeTake & \ffRawPtrFnsSafeTake & \fttRawPtrFnsSafeTake & \nextRawPtrFnsSafeTake \\
    Types with a hand-written \texttt{Drop} & \lgTyDrop{} (\lgTyDropPct\%) & \gitTyDrop{} (\gitTyDropPct\%) & \ffTyDrop{} (\ffTyDropPct\%) & \fttTyDrop{} (\fttTyDropPct\%) & \nextTyDrop{} (\nextTyDropPct\%) \\
    Types with a reference-forming trait & \lgTyRefForming & \gitTyRefForming{} (\gitTyRefFormingPct\%) & \ffTyRefForming & \fttTyRefForming{} (\fttTyRefFormingPct\%) & \nextTyRefForming{} (\nextTyRefFormingPct\%) \\
    Types storing a raw pointer & \lgTyRawField{} (\lgTyRawFieldPct\%) & \gitTyRawField{} (\gitTyRawFieldPct\%) & \ffTyRawField{} (\ffTyRawFieldPct\%) & \fttTyRawField{} (\fttTyRawFieldPct\%) & \nextTyRawField{} (\nextTyRawFieldPct\%) \\
    \bottomrule
  \end{tabular}
\end{table}

\subsection{RQ1: Safety and Ergonomics}
\label{sec:eval-guarantees}
\paragraph{Unsafe surface.}
Of the bindings' \lgPubFns{} and \ffPubFns{} public functions, counting the methods of their trait implementations, \lgUnsafeFnPub{} and \ffUnsafeFnPub{} are \texttt{unsafe}, most of them not by the bindings' choice: \lgUnsafeFnTrait{} and \ffUnsafeFnTrait{} implement \texttt{ffibox}'s lifecycle policies, e.g., \texttt{CDrop::c\_drop}, which \texttt{ffibox}'s traits declare \texttt{unsafe} and only \texttt{ffibox}'s owners call, and \lgUnsafeFnSeam{} and \ffUnsafeFnSeam{} are seams, conversions that the type system does not check, across the boundary or between Rust types, e.g., \texttt{from\_raw}. The remaining \lgUnsafeFnOther{} and \ffUnsafeFnOther{} (\lgUnsafeFnOtherPct\% and \ffUnsafeFnOtherPct\% of the public functions) are \texttt{unsafe} by choice, mostly for obligations that no Rust type expresses. In libgit2, \lgUnsafeFnGlobal{} set or read libgit2's process-wide options and trace callback, which libgit2 does not synchronize and git2-rs also declares \texttt{unsafe}; \lgUnsafeFnBackend{} install backends implemented in C, e.g., \texttt{RepositoryRef::set\_refdb}; and \lgUnsafeFnOdb{} clear or write an object database's commit graph and multi-pack index, which no other thread may use meanwhile through any of the database's owners. In ffmpeg, \ffUnsafeFnStd{} mirror \texttt{unsafe} APIs of the standard library and memmap2, \texttt{MaybeUninit::assume\_init} and \texttt{Mmap::map}, and \ffUnsafeFnCRec{} rely on what only C records, such as the allocation behind a frame's plane pointers or a buffer pool's allocator. The other \lgUnsafeFnLift{} and \ffUnsafeFnLift{} could be made safe, e.g., \texttt{cpu::force\_cpu\_flags} by checking the flags against those the CPU reports. Outside seams and \texttt{ffibox}'s traits, \lgRawPtrFns{} and \ffRawPtrFns{} public functions (\lgRawPtrFnsPct\% and \ffRawPtrFnsPct\%) have a raw pointer in their signature: in libgit2, both take a pointer to a C type that the bindings do not wrap: \texttt{Refdb::new} a reference-database backend, which a later wave of its own module wraps (\S\ref{sec:wavefront}), and \texttt{add\_ssl\_x509\_cert} an OpenSSL certificate, which another library owns; in ffmpeg, \ffRawPtrFnsOpaque{} store or return opaque user data, and two set a frame's plane pointers. \lgRawPtrFnsSafeTake{} and \ffRawPtrFnsSafeTake{} of them are safe functions that take one. Over the \lgPubWrapFns{} and \ffPubWrapFns{} public functions that wrap a C function, \texttt{ffibox}'s policies aside, \lgUnsafeWrapFns{} and \ffUnsafeWrapFns{} (\lgUnsafeWrapFnsPct\% and \ffUnsafeWrapFnsPct\%) are \texttt{unsafe} and \lgRawPtrWrapFns{} and \ffRawPtrWrapFns{} (\lgRawPtrWrapFnsPct\% and \ffRawPtrWrapFnsPct\%) have a raw pointer in their signature; libgit2's share of \texttt{unsafe} wrappers is higher than over all its public functions because every one of its \lgUnsafeFnGlobal{} process-wide option and trace functions wraps a C function. 

Counting every public function, our libgit2 bindings declare a slightly smaller share \texttt{unsafe} than git2-rs (\lgUnsafeFnPct\% against \gitUnsafeFnPct\%), and our ffmpeg bindings a smaller one than ffmpeg-the-third and ffmpeg-next (\ffUnsafeFnPct\%, against \fttUnsafeFnPct\% and \nextUnsafeFnPct\%), but most of the community crates' \texttt{unsafe} functions are seams that hand a C pointer to or from the consumer, such as git2-rs's \texttt{Binding::from\_raw}, and few are \texttt{unsafe} by choice in any of the five (Table~\ref{tab:baselines}). A raw pointer in the signature of a safe function is sound when the function returns it, but not when it adopts a pointer that safe code later dereferences: neither libgit2 crate has such a function, and the \ffRawPtrFnsSafeTake{} in our ffmpeg bindings store user-data pointers that neither ffmpeg nor the bindings dereference, whereas among the \nextRawPtrFnsSafeTake{} of ffmpeg-next are safe constructors such as \texttt{codec::video::RateIter::new}, whose iterator dereferences the pointer, so that safe code can cause \gls{ub} by passing any address. Behind this interface, \texttt{unsafe} blocks account for \lgUnsafeLocPct\% and \ffUnsafeLocPct\% of the bindings' non-test code. This surface covers only what the bindings wrap, however: a consumer that calls one of the \lgUnwrappedFns{} and \ffUnwrappedFns{} public C functions that remain unwrapped passes it a wrapped object through a seam and an \texttt{unsafe} call. Wrapped types sit behind a pointer in \lgSeamPos{} and \ffSeamPos{} parameters and returns of these functions, most often libgit2's \texttt{git\_repository} (\lgSeamPosRepo{} parameters) and ffmpeg's \texttt{AVFrame} (\ffSeamPosFrame{}), and in \lgSeamFld{} and \ffSeamFld{} fields of unwrapped structs; another \lgValPos{} and \ffValPos{} parameters and returns hold one by value, which needs only a conversion.

\paragraph{\texttt{ffibox} usage.}
Through the rustdoc aliases of \S\ref{sec:scan} (Table~\ref{tab:ffibox-use}), the bindings wrap \lgSysFnsWrappedPct\% and \ffSysFnsWrappedPct\% of the C functions that \texttt{libgit2-sys} and \texttt{libavutil-sys} declare; the functions they leave unwrapped serve the tests, which call C directly, and the ffmpeg bindings wrap \ffSysFnsWrappedNotUsed{} C functions that they reimplement rather than call, such as the \texttt{static inline} \texttt{av\_uuid\_copy}. Of the \texttt{-sys} structs that the bindings use, \lgSysTysWrappedPct\% and \ffSysTysWrappedPct\% are wrapped with a layout type, i.e., a type implementing \texttt{CCell}.

\texttt{ffibox} carries most of the bindings' types: \lgTyFfiboxPct\% and \ffTyFfiboxPct\% of their public types implement one of its traits, as layout types and their handles or as lifecycle policies, and most of the others are plain Rust values, such as enums of C constants and error types. The \lgCtypeManual{} libgit2 layout types that \texttt{define\_ctype!} did not emit carry lifetimes, which the macro does not support (\S\ref{sec:synthesis}). Public signatures return \texttt{ffibox} owners in \lgSigOwnerRet{} and \ffSigOwnerRet{} positions and take them in \lgSigOwnerArg{} and \ffSigOwnerArg{}, and a layout type that a signature names by value is a value type whose bytes Rust owns, such as \texttt{Oid} and \texttt{Rational}. Field accessors are methods of a handle rather than of a layout type: \lgFieldAccRef{} and \ffFieldAccRef{} read a field through a shared handle and \lgFieldAccMut{} and \ffFieldAccMut{} write one through an exclusive handle, so no accessor forms a reference to the C object; the other \lgFieldAccOther{} and \ffFieldAccOther{} stand for a C vtable's \texttt{free} member, which a policy implements, for an operation that consumes its owner, or for storage that Rust owns. Only immutable globals occur in our campaigns, which the bindings expose as plain shared references, and no callback was forked by contract.

\paragraph{Outside \texttt{ffibox}.}
The bindings bypass \texttt{ffibox}'s primitives in a few places. No standard owner (e.g., \texttt{Box} or \texttt{Arc}) holds a layout type, and no type implements a reference-forming trait such as \texttt{Deref}. Only \lgTyDrop{} and \ffTyDrop{} public types (\lgTyDropPct\% and \ffTyDropPct\%) implement \texttt{Drop} by hand: guards over C state rather than C objects, such as libgit2's \texttt{Runtime}, which shuts libgit2 down; owners of Rust storage that C reads, such as a write stream implemented in Rust; and collections that \texttt{CVec} cannot represent, a vector whose strings own memory and, in ffmpeg, arrays that C grows in place through a pointer and an \texttt{int} count lent by address. The community crates, by contrast, release C objects in hand-written \texttt{Drop} implementations in \gitTyDropPct\%, \fttTyDropPct\%, and \nextTyDropPct\% of their public types, and implement reference-forming traits in up to \nextTyRefFormingPct\%. A type built on \texttt{ffibox} declares no raw pointer of its own: the pointer of the C object it manages sits inside a \texttt{CBox}, \texttt{CArc}, or \texttt{CVec} whose trait declares its lifecycle, or inside the C struct that a layout type covers. Counting the raw pointers, \texttt{NonNull}s, and \texttt{AtomicPtr}s that a crate declares in the fields of the structs it declares itself, \lgTyRawField{} and \ffTyRawField{} of our public types hold one (\lgTyRawFieldPct\% and \ffTyRawFieldPct\%), against \gitTyRawFieldPct\%, \fttTyRawFieldPct\%, and \nextTyRawFieldPct\% of the community crates' (Table~\ref{tab:baselines}). Ours are, in libgit2, a view that borrows Rust strings, a write stream implemented in Rust, and a configuration identity that is never dereferenced, and in ffmpeg, C memory that no layout type covers: handles to a type-erased object, i.e., any C struct whose first member points to an \texttt{AVClass}, C's iteration state, the in-place arrays and their views, a log record's \texttt{va\_list}, and two blocks whose layout C keeps to itself. The C structs used without a layout type are libgit2's, which the bindings reach before the waves that wrap them, and ffmpeg's \texttt{va\_list}. Outside the sanctioned receivers, public signatures and fields hold \lgRefRest{} and \ffRefRest{} references to a layout type: \lgRefRestCmp{} and \ffRefRestCmp{} are the second operand of a comparison, which the standard traits require, such as \texttt{PartialEq::eq(\&self, other: \&Self)}, and most of the others are slices of value types that Rust holds, such as \texttt{\&[Oid]}; none refers to memory that C holds, and these counts are lower bounds (\S\ref{sec:scan}).

\begin{table}
  \caption{\texttt{ffibox} use in the bindings, over the C functions and structs of the wrapped library's \texttt{-sys} crate.}
  \label{tab:ffibox-use}
  \small
    \begin{threeparttable}
  \begin{tabular}{@{}lrr@{}}
    \toprule
     & libgit2 & ffmpeg \\
    \midrule
    C functions wrapped / declared & \lgSysFnsWrapped{} / \lgSysFns & \ffSysFnsWrapped{} / \ffSysFns \\
    Used C structs with a layout type & \lgSysTysWrapped{} / \lgSysTysUsed & \ffSysTysWrapped{} / \ffSysTysUsed \\
    C fields with an accessor & \lgFields & \ffFields \\
    \quad accessors, on shared / exclusive handles & \lgFieldAccRef{} / \lgFieldAccMut & \ffFieldAccRef{} / \ffFieldAccMut \\
    Public types on \texttt{ffibox} (\%) & \lgTyFfiboxPct & \ffTyFfiboxPct \\
    Layout types: \texttt{define\_ctype!} / by hand & \lgCtypeMacro{} / \lgCtypeManual & \ffCtypeMacro{} / \ffCtypeManual \\
    \texttt{ffibox} owners returned / taken\tnote{a} & \lgSigOwnerRet{} / \lgSigOwnerArg & \ffSigOwnerRet{} / \ffSigOwnerArg \\
    \midrule
    Hand-written \texttt{Drop} / raw-pointer types & \lgTyDrop{} / \lgTyRawNoFfibox & \ffTyDrop{} / \ffTyRawNoFfibox \\
    Used C structs without a layout type & \lgSysTysRaw & \ffSysTysRaw \\
    Unsanctioned references to layout types\tnote{a} & \lgRefRest & \ffRefRest \\
    \bottomrule
  \end{tabular}
  \begin{tablenotes}[flushleft]\footnotesize
  \item[a] Positions in public signatures and fields.
  \end{tablenotes}
  \end{threeparttable}
\end{table}

\subsection{RQ2: Soundness Evidence}
\label{sec:eval-defects}

\paragraph{Tests.}
Sanitizers and Miri detect \gls{ub} only on the paths that tests execute, so the bindings' tests and their coverage bound the evidence that the sanitizers provide. The bindings carry \lgTests{} and \ffTests{} tests in the four suites of \S\ref{sec:testing} (Table~\ref{tab:coverage}), which together cover \lgCovRust\% and \ffCovRust\% of the bindings' non-test Rust code, and \lgCovC\% of libgit2's C code and \ffCovC\% of libavutil's, the only ffmpeg library wrapped so far. Both languages are measured with LLVM's source-based coverage from one toolchain, over the code under test only. At both campaign tips, every suite passes on the plain build and under \gls{asan}, \gls{ubsan}, and \gls{tsan}, and the workspace passes BorrowSanitizer, which skips the compile-fail suite and, in the libgit2 bindings, \lgBsanSkip{} tests: \lgBsanSkipSlow{} whose inputs outlast the gate, \lgBsanSkipHang{} that hangs on a child process's abort path, and \lgBsanSkipFp{} that would trigger BorrowSanitizer false positives. The community crates' suites run no sanitizers and are much smaller: the generated bindings carry roughly an order of magnitude more tests than git2-rs and more than fifty times as many as the ffmpeg crates, and cover nearly all of their Rust code, against 61\% for git2-rs and under 20\% for the ffmpeg crates. On the C side, git2-rs's suite reaches about as much of libgit2 as ours (\gitCovC\% against \lgCovCCmp\%, both over libgit2 with its bundled xdiff and llhttp), whereas the ffmpeg crates reach about a tenth of libavutil at most, against two thirds for our bindings (Table~\ref{tab:baselines}). Since Miri cannot execute foreign code, it runs only the \lgMiriRun{} and \ffMiriRun{} tests that make no \gls{ffi} call (\lgMiriRunPct\% and \ffMiriRunPct\% of the suites), all of which pass.

\begin{table}
  \caption{Line coverage per test suite at the campaign tips, over the bindings' non-test sources (Rust) and over libgit2 and libavutil, the only ffmpeg library wrapped so far (C). \emph{All} is the union of all suites.}
  \label{tab:coverage}
  \small
  \begin{tabular}{@{}lrrrrrr@{}}
    \toprule
    & \multicolumn{3}{c}{libgit2} & \multicolumn{3}{c}{ffmpeg} \\
    \cmidrule(lr){2-4}\cmidrule(l){5-7}
    Suite & Tests & Rust (\%) & C (\%) & Tests & Rust (\%) & C (\%) \\
    \midrule
    Safe UB & \lgTestsUbSafe & \lgUbCovRust & \lgUbCovC & \ffTestsUbSafe & \ffUbCovRust & \ffUbCovC \\
    Unsafe UB & \lgTestsUbUnsafe & \lgUbUnsafeCovRust & \lgUbUnsafeCovC & \ffTestsUbUnsafe & \ffUbUnsafeCovRust & \ffUbUnsafeCovC \\
    Equivalence & \lgTestsEquiv & \lgEqCovRust & \lgEqCovC & \ffTestsEquiv & \ffEqCovRust & \ffEqCovC \\
    Unit & \lgTestsUnit & \lgUnitCovRust & \lgUnitCovC & \ffTestsUnit & \ffUnitCovRust & \ffUnitCovC \\
    \midrule
    All & \lgTests & \lgCovRust & \lgCovC & \ffTests & \ffCovRust & \ffCovC \\
    \bottomrule
  \end{tabular}
\end{table}

\paragraph{Agentic review.}
The review waves filed \lgReportsAll{} and \ffReportsAll{} reports (Table~\ref{tab:findings}), each with a fix; \gls{ub} and equivalence defects also come with a reproducer. Three categories bear directly on safety. First, \lgReportsUbSafe{} and \ffReportsUbSafe{} defects are \gls{ub} reachable from safe Rust, and \lgReportsUbUnsafe{} and \ffReportsUbUnsafe{} more are \gls{ub} reachable through the \texttt{unsafe} API within its contract. By kind, they comprise \lgUbOob{} and \ffUbOob{} out-of-bounds accesses, \lgUbUaf{} and \ffUbUaf{} uses after free, \ffUbDf{} double free in ffmpeg, \lgUbNull{} null-pointer dereference in libgit2, and no aliasing violation; the others are \ffUbArith{} arithmetic overflows in ffmpeg's C, a data race, and a misaligned slice. For instance, inserting a SHA-256 object ID into a SHA-1 repository's tree builder through libgit2's safe API made C write past a heap buffer, which the wrapper now refuses.

Second, \lgReportsUnsafe{} and \ffReportsUnsafe{} reports concern the \texttt{unsafe} boundary: \lgRvSurface{} and \ffRvSurface{} made the public surface safer, removing \lgRvUnsafeFnRemoved{} and \ffRvUnsafeFnRemoved{} public \texttt{unsafe} functions and, in ffmpeg, \ffRvRawPtrRemoved{} raw pointers, or closing unsound safe entry points, such as libgit2's error setters, which now take the typed \texttt{ErrorClass} instead of a raw integer that safe code could set to any value, and \lgRvInternal{} and \ffRvInternal{} replaced internal raw-pointer code with safe helpers. Third, \lgReportsObligations{} and \ffReportsObligations{} corrected a \texttt{/// \# Safety} obligation or a \texttt{// SAFETY:} justification, \texttt{Send} and \texttt{Sync} ones included. Review thus changed the public surface (\S\ref{sec:eval-guarantees}) no more often than it corrected \gls{ub} or safety obligations, since that surface was already small when the binders landed it. In addition, \lgReportsEquiv{} and \ffReportsEquiv{} reports are behavioral divergences from C. Most remaining reports concern coding conventions and miscellaneous fixes to documentation and tests. Review fills the coverage gaps it finds with tests rather than reports, and its fixes come with regression tests: review batches added \lgRvAddTests{} and \ffRvAddTests{} tests, \lgRvAddTestsPct\% and \ffRvAddTestsPct\% of the suites, which raise the suites' coverage by \lgRvAddCovRust{} and \ffRvAddCovRust{} percentage points on the Rust side and by \lgRvAddCovC{} and \ffRvAddCovC{} points on the C side, and a single suite by up to \lgRvAddUnitCovRust{} points, since the binding batches' own tests already reach nearly all of the bindings.

\paragraph{Review of a community crate.} We also ran Crustify's reviewers, with the \gls{ub}, contract, and equivalence objectives and the same instruments, on git2-rs~\gitAuditVersion{}. In about \gitAuditWallHours{} hours of wall-clock time and \$\gitAuditCostUSD{}, covering \gitAuditFiles{} of its \gitAuditFilesAll{} source files, they found \gitAuditUb{} defects that cause \gls{ub} reachable from safe code, each confirmed by a reproducer: \gitAuditUbAsan{} under \gls{asan} or \gls{ubsan}, \gitAuditUbTsan{} under \gls{tsan}, \gitAuditUbBsan{} under BorrowSanitizer, and \gitAuditUbMiri{} under Miri, none of which git2-rs's \gls{ci} can detect, since it runs neither sanitizers nor Miri. We are disclosing them to its maintainers and withhold their details until they are fixed.

\begin{table}
  \caption{Review reports per category. \emph{UB (safe)} and \emph{UB (unsafe)} are reachable from the safe API and from the \texttt{unsafe} API within its contract.}
  \label{tab:findings}
  \small
    \begin{threeparttable}
  \begin{tabular}{@{}lrr@{}}
    \toprule
    Category & libgit2 & ffmpeg \\
    \midrule
    UB (safe) & \lgReportsUbSafe{} & \ffReportsUbSafe{} \\
    UB (unsafe) & \lgReportsUbUnsafe{} & \ffReportsUbUnsafe{} \\
    Residual \texttt{unsafe} or raw types & \lgReportsUnsafe{} & \ffReportsUnsafe{} \\
    Safety obligations & \lgReportsObligations{} & \ffReportsObligations{} \\
    Equivalence & \lgReportsEquiv{} & \ffReportsEquiv{} \\
    Other\tnote{a} & \lgReportsOther{} & \ffReportsOther{} \\
    \midrule
    All reports & \lgReportsAll{} & \ffReportsAll{} \\
    \bottomrule
  \end{tabular}
  \begin{tablenotes}[flushleft]\footnotesize
  \item[a] Missing lifecycle routines or owners, internal defects, deviations from the coding conventions, miscellaneous fixes, and reports on the ownership and lifecycle records that agents submitted to Wavefront (\lgReportsWavefront{} and \ffReportsWavefront{}).
  \end{tablenotes}
  \end{threeparttable}
\end{table}

\subsection{RQ3: Coverage and Cost}
\label{sec:eval-cost}
\label{sec:eval-coverage}
\paragraph{API coverage.}
Crustify and one developer wrapped \lgWrappedItems{} of libgit2's \lgApiItems{} public items (\lgApiCov\%), i.e., \lgWrappedTypes{} of its \lgApiTypes{} types (\lgApiCovTypes\%) and \lgWrappedSyms{} of its \lgApiSyms{} symbols (\lgApiCovSyms\%), and \ffWrappedItems{} of ffmpeg's \ffApiItems{} public items (\ffApiCov\%, counted over ffmpeg as a whole), i.e., \ffWrappedTypes{} of its \ffApiTypes{} types (\ffApiCovTypes\%) and \ffWrappedSyms{} of its \ffApiSyms{} symbols (\ffApiCovSyms\%). Its binding batches inserted \lgLocInserted{} and \ffLocInserted{} lines of Rust, tests included, of which \lgCodeLines{} (\lgCodeLinesPct\%) and \ffCodeLines{} (\ffCodeLinesPct\%) are non-test code. We stopped both campaigns before they covered the entire API mainly to spare tokens and share our findings. Even so, the ffmpeg bindings cover more than twice as much of the public API as either community crate, even at their upper bounds, whereas git2-rs's upper bound, \gitApiCov\%, exceeds our \lgApiCov\%, so git2-rs may cover more of libgit2 (Table~\ref{tab:baselines}).

\paragraph{Cost.}
The libgit2 campaign used about \lgTokensB{} billion tokens (\$\lgCostUSD{}) and the ffmpeg campaign so far about \ffTokensB{} billion tokens (\$\ffCostUSD{}). Agents ran for \lgWallHours{} and \ffWallHours{} hours of wall-clock time, running the batches of each wave concurrently (\S\ref{sec:eval-wavefront}); run one batch at a time, they would have taken \lgAgentHoursSum{} and \ffAgentHoursSum{} hours, i.e., \lgSeqSpeedup{} and \ffSeqSpeedup{} times as long. The planners, binders, and reviewers account for \$\lgAgentsCostUSD{} and \$\ffAgentsCostUSD{} of the cost, and the assistant for \$\lgOrchCostUSD{} and at least \$\ffOrchCostUSD{}: \$\lgAsstOrchUSD{} and \$\ffAsstOrchUSD{} to orchestrate the campaigns, \$\lgAsstPlanUSD{} and \$\ffAsstPlanUSD{} to plan them, and \$\lgAsstDevUSD{} and \$\ffAsstDevUSD{} to answer the developer's queries. A type cost \$\usdPerType{} to wrap on average, between \$\usdPerTypeMin{} and \$\usdPerTypeMax{}, largely independently of the number of fields it wraps, whereas the cost of a symbol grows with the C code it wraps, at \$\usdPerSym{} per symbol or about \$\usdPerKloc{} per thousand lines of C; review accounts for \lgReviewShare\% and \ffReviewShare\% of the agents' cost (Appendix~\ref{sec:appendix-batches}).

\subsection{RQ4: Dependency-Ordered Scheduling}
\label{sec:eval-wavefront}

\paragraph{Graphs and schedules.} Both campaigns are scheduled from public-signature graphs of 7{,}526 and 5{,}651 nodes. Wavefront orders the campaigns' 95 and 160 modules into 9 and 10 module layers and schedules their 1{,}144 and 1{,}337 items, i.e., the public APIs of \S\ref{sec:eval-cost} but for 2 and 1 macros, which the bindings reach through the items that use them, in 204 and 281 waves of up to 5 and 9 batches each; since independent modules run concurrently, the longest chain of dependent waves spans 10 and 12 waves. Of these, 67 and 98 waves hold the items that the campaigns have landed so far, under an earlier schedule. Cycles are rare: 29 and 11 of the nodes (0.39\% and 0.19\%) lie on a cycle, and Wavefront cuts 22 and 6 edges, i.e., 0.30\% and 0.18\% of all edges, to break them, and 22 and 7 dependencies between modules. In the implementation graphs, which also follow function bodies, 3.4\% and 1.5\% of the nodes lie on cycles, the largest of which spans 135 nodes and 5{,}147 lines of C in libgit2. Pipelines that condense cycles into single units (\S\ref{sec:related}) would translate this cycle as one unit, whereas Wavefront's unit is always a single item. Wavefront's one-time analysis of each target took about 17 and 24 minutes (Appendix~\ref{sec:appendix-wavefront}).

\paragraph{Concurrency.} The campaigns landed these items in \lgExecWaves{} and \ffExecWaves{} waves, \lgMultiModWaves{} and \ffMultiModWaves{} of which ran batches of several modules concurrently. A wave ran between 1 and \lgWaveBatchesMax{} and between 1 and \ffWaveBatchesMax{} batches at once, \lgWaveBatchesMedian{} and \ffWaveBatchesMedian{} at the median, and up to \lgPeakBatches{} batches ran at once across the campaign. Since waves ran one after the other, a single batch ran for \lgSoloPct\% and \ffSoloPct\% of the time; \S\ref{sec:eval-cost} reports the wall-clock time that this concurrency saved.

\paragraph{Corrections.} The planners corrected the relations that casts and type-erased slots hide (D2) in \lgTthreeRows{} and \ffTthreeRows{} rows of the campaigns' current plans: \lgTthreeAdd{} and \ffTthreeAdd{} add a relation that the extraction missed, i.e., a type that a \texttt{void *} parameter is read as or a base--derived pair, and \lgTthreeDrop{} and \ffTthreeDrop{} drop a spurious one, e.g., a slot that one file reuses for an unrelated type. In ffmpeg, all \ffTthreeErasedArg{} added parameter rows resolve a logging or option context, a \texttt{void *} to any struct whose first member points to an \texttt{AVClass}: the first waves, planned without these rows, bound such contexts as \texttt{void *} in \texttt{unsafe} functions, and the re-planned waves bind them as typed handles, e.g., \texttt{LogContext}. libgit2's planners also named \lgTthreeGenMacros{} macros as the generators of \lgTthreeGenRows{} types, such as its \texttt{GIT\_HASHMAP\_STRUCT} family. Each row was reviewed twice: by the later planner runs, which re-check what earlier ones decided (\S\ref{sec:agents}), and by the developer before approving the plan. The later runs kept \lgTthreeKept{} of the first run's \lgTthreeFirst{} rows in libgit2 and added \lgTthreeLater{}, and kept \ffTthreeKept{} of \ffTthreeFirst{} in ffmpeg, dropped one, and added \ffTthreeLater{}, each with a report.

\subsection{RQ5: Ownership and Lifecycle Reasoning}
\label{sec:eval-ownership}

\paragraph{Pointer contracts.}
The agents recorded the ownership of \lgOwnPos{} and \ffOwnPos{} pointers in fields, parameters, and returns: \lgOwnOwned{} and \ffOwnOwned{} own their pointee and \lgOwnBorrowed{} and \ffOwnBorrowed{} borrow it, \lgOwnBorrowArgPct\% and \ffOwnBorrowArgPct\% of the latter for another argument's lifetime; \lgOwnMutablePct\% and \ffOwnMutablePct\% are mutable and \lgOwnNullablePct\% and \ffOwnNullablePct\% nullable, and \lgOwnString{} and \ffOwnString{} point to strings and \lgOwnArray{} and \ffOwnArray{} to other arrays.

\paragraph{Lifecycle contracts.}
The agents also identified \lgOwnCtors{} and \ffOwnCtors{} constructors, \lgOwnDroppers{} and \ffOwnDroppers{} destructors, \lgOwnCloners{} and \ffOwnCloners{} cloners, \lgOwnRefcount{} and \ffOwnRefcount{} reference counts, and \lgOwnLocked{} and \ffOwnLocked{} lock-protected fields.

\paragraph{Review of the contracts.}
Reviewers re-check the records of every item they review (\S\ref{sec:review}), and every landed wave was reviewed. They filed \lgReportsWavefront{} and \ffReportsWavefront{} reports on the records (Table~\ref{tab:findings}), each with the evidence, which we classified against the changes they made to the ownership store. Review found \lgOwnFp{} and \ffOwnFp{} false positives, i.e., wrong pointer records (\lgOwnFpPct\% and \ffOwnFpPct\% of the positions), such as a nullable out-parameter recorded as non-null or a retained borrow recorded as lasting only the call, and no wrong lifecycle role. It found more false negatives: \lgOwnFnPos{} and \ffOwnFnPos{} positions (\lgOwnFnPosPct\% and \ffOwnFnPosPct\%) of functions that binders had not recorded, and \lgOwnFnLife{} and \ffOwnFnLife{} lifecycle roles (\lgOwnFnLifePct\% and \ffOwnFnLifePct\%), mostly constructors and cloners. The other reports refine the records' notes, e.g., the conditions under which a call is UB. These counts bound the agents' errors on the landed waves from below: review is not exhaustive, so they do not establish that the remaining records are correct. Appendix~\ref{sec:appendix-ownership} compares the records with static ownership inference.

\section{Discussion}
\label{sec:discussion}

\paragraph{Threats to validity.}
The models we used were likely trained on the community-maintained bindings that we compare against, and may draw on their design. They cannot copy them, however: Crustify's bindings are built from \texttt{ffibox}, which did not exist before this work, and \lgTyFfiboxPct\% and \ffTyFfiboxPct\% of their public types implement \texttt{ffibox}'s traits (Table~\ref{tab:ffibox-use}). Moreover, the ffmpeg bindings cover more than twice as much of the public API as either community crate, so at least half of the items they wrap have no counterpart in either crate; for libgit2, this argument applies only to the representation. Crustify's prompts and gates changed while the campaigns ran, and one developer supervised both (\S\ref{sec:eval-setup}): earlier waves thus ran under less strict rules than later ones, and since we did not run a campaign without the developer, we do not measure how much their checks contributed. Each batch of the reported campaigns ran once. Since \gls{llm} agents are not deterministic, repeated runs of the same batch may produce different bindings, tests, and review findings. We therefore report the reproducible artifacts of one campaign per target rather than averages over runs. Finally, Crustify's safety and soundness evidence is empirical and therefore incomplete. Sanitizers and Miri detect \gls{ub} only on the paths that tests execute, so the absence of reports is bounded by the test coverage that we measure; ffimap is a linter rather than a verifier (\S\ref{sec:scan}), so its counts of references into C memory and of primitives written outside \texttt{ffibox} may miss sites; and review finds defects but does not show their absence. The defects that we report are thus a lower bound on the defects in the bindings.

\paragraph{Future work.}
Safe bindings are also needed when a production codebase is ported to Rust incrementally, subsystem by subsystem: until the port completes, many pointers cross the \gls{ffi} boundary between ported and unported subsystems, in both directions. Since Crustify already decomposes a target into modules and schedules them in dependency order, we plan to extend it to porting, with safe bindings at each boundary between ported and unported code. C++ libraries have safe bindings too, e.g., rusty\_v8~\cite{rustyv8} for V8, and we plan to support C++ targets next. C++ carries ownership and lifecycles in smart pointers and constructors, but its borrowed pointers and references still carry no lifetime, so their contracts must be inferred as for C. Finally, \texttt{ffibox} does not yet represent two collections that the bindings implement by hand (\S\ref{sec:eval-guarantees}): vectors whose elements own memory, and arrays that C grows in place through a pointer and a count lent by address.

\section{Related Work}
\label{sec:related}

\paragraph{FFI primitives.} Hand-written binding crates, e.g., git2-rs~\cite{git2rs}, rusqlite~\cite{rusqlite}, curl-rust~\cite{curlrust}, and ffmpeg-next~\cite{ffmpegnext}, encode a C library's ownership conventions type by type, with hand-written \texttt{Drop} implementations and references into C-owned memory, in ffmpeg-next even \texttt{\&mut [u8]} over frame buffers that C may share. aws-lc-rs~\cite{awslcrs} factors its owners into a generic pointer type, and ffmpeg-the-third~\cite{ffmpegthethird} adds lifetime-bound \texttt{Ref} and \texttt{Mut} handles for some types, but both remain hand-written per type. foreign-types~\cite{foreigntypes}, used by rust-openssl, fixes one destructor per type with a macro and hands out borrowed objects as \texttt{\&FooRef}, and Rust for Linux~\cite{rflkernel,li2024rfl} exposes kernel objects as \texttt{\&T} over an \texttt{UnsafeCell}-based \texttt{Opaque<T>}; such references can cause \gls{ub} at \gls{ffi} boundaries~\cite{mccormack2025ffi}. Closest to \texttt{ffibox}, Omniglot~\cite{schuermann2025omniglot,schuermann2023ef} hands out a reference into foreign memory only after run-time validation and until the next foreign call, but targets untrusted libraries, pays for isolation at run time, and does not track the lifecycle of individual foreign objects. Systems that isolate foreign code~\cite{lamowski2017sandcrust,rivera2021galeed,kirth2022pkrusafe} or detect cross-language memory-management errors~\cite{li2022ffichecker} do not let safe Rust own or borrow C objects in place. \texttt{ffibox} instead never forms a reference to C-owned memory, names lifecycle routines through policy types, and withholds \texttt{Send} and \texttt{Sync} unless a binding argues for them, at no run-time cost.

\paragraph{Type-dependency ordering.} Repository-level translation pipelines order their units over call graphs~\cite{nitin2026c2saferrust}, types before methods~\cite{alphatrans2025}, or one graph of types, globals, and functions~\cite{sactor2026,cornelius2026ideas,ptrtrans2026}, as Wavefront does. On cycles, RustMap~\cite{rustmap2025}, IDEAS, and PtrTrans condense each \gls{scc} into one translation unit, SACTOR rejects cyclic inputs, and C2SaferRust and AlphaTrans break cycles where their traversal does. Wavefront instead splits cycles with a weighted form of the vertex ordering of Eades et al.'s feedback-arc-set heuristic~\cite{eades1993fas} and layers the graph by longest path~\cite{sugiyama1981}, so that a wave's batches can be wrapped concurrently, and it keeps lifecycle routines as nodes of their own, whereas PtrTrans removes deallocation functions from its graph. Indexers such as clangd, Joern, and CodeQL~\cite{clangd,joern2014cpg,ql2016} expose the facts of such a graph, not the graph.

\paragraph{Ownership inference.} Ownership inference for C pointers mainly serves translation into safe Rust. Laertes~\cite{emre2021laertes} rewrites raw pointers in C2Rust output using feedback from the borrow checker; Crown~\cite{zhang2023crown} infers ownership with a SAT-based analysis, but generates no constraints when a library wraps its allocator behind function pointers; and PtrTrans~\cite{ptrtrans2026} derives annotations from fixed rules that label each pointer as owning or borrowing and every pointer as nullable. \gls{llm}-based approaches infer the Checked C bounds of pointers~\cite{mohammed2025msa} or the functions that allocate or free memory~\cite{shemetova2025lamed}, but not borrows, conditional ownership, cloners, or reference counts. Closest to Crustify, \&inator~\cite{chen2026inator} infers an ownership and borrowing discipline for a C program's interface with a whole-program SMT encoding, but chooses one Rust type per pointer for a rewrite into Rust-owned memory rather than for bindings over unchanged C, and does not yet support function pointers, unions, or multi-threaded programs (Appendix~\ref{sec:appendix-ownership}).

\paragraph{Safety measurement.} cargo-geiger~\cite{cargogeiger}, the unsafe metric of PtrTrans, counts \texttt{unsafe} items and expressions with a parser, and cargo-scan~\cite{zoghbi2026cargoscan} reports effects, such as \gls{ffi} calls, that reach a crate's \texttt{pub} functions; Astrauskas et al.~\cite{astrauskas2020unsafe} note that \texttt{pub} does not distinguish visibility within a crate from visibility to other crates. C-to-Rust translators count raw pointers or unsafe lines or tokens over the whole program~\cite{emre2021laertes,zhang2023crown,nitin2026c2saferrust,smartc2rust2026,sactor2026,ling2022crusts}. ffimap instead counts what a binding exposes to its consumers. Unlike either tool, it also classifies each raw pointer by its position, visibility, and the safety of its function, and treats the types imported from a \texttt{-sys} crate as shared with C: it finds the raw pointers that a binding's types hold through them, and the C structs that a binding uses raw rather than through a layout type (Appendix~\ref{sec:appendix-ffimap}).

\paragraph{Multi-agent systems for C-to-Rust tasks.} Most \gls{llm}-based, project-level C-to-Rust systems drive one model through a fixed pipeline of translation and compiler-guided repair~\cite{nitin2026c2saferrust,smartc2rust2026,rustmap2025,ptrtrans2026,evoc2rust2026,sactor2026}. ReCodeAgent~\cite{recodeagent2026} runs its analyzer, planner, translator, and validator agents as separate sessions, and Schesch and Ernst~\cite{schesch2026agentic} let a coding agent delegate subtasks to subagents; neither validates with a model other than the translator's, and both process units one at a time, whereas Crustify's reviewers mostly ran on a different model than the binders (\S\ref{sec:eval-setup}) and a wave's batches run concurrently. None of these systems generates safe bindings to unchanged C: \gls{ffi} serves them only as a verification scaffold~\cite{sactor2026} or as an exported C interface~\cite{schesch2026agentic}. Apart from Schesch and Ernst's equivalence tests, their validation relies mostly on existing tests (\S\ref{sec:intro}), and none measures line coverage of the C and the Rust code in the same \gls{ffi}-linked run to guide test generation, or feeds sanitizer and Miri reports back to its agents.

\section{Conclusion}
\label{sec:conclusion}

We presented Crustify, an agentic co-translator that generates safe Rust bindings for repository-level C libraries, designed to be sound and ergonomic. Its agents bind dependency-ordered batches concurrently with \texttt{ffibox}'s primitives, derive the ownership and lifecycle contracts of the code they wrap, which Wavefront stores, and check the bindings with coverage-guided tests under sanitizers and Miri and with independent review, while ffimap maps their \texttt{unsafe} surface. On libgit2 and ffmpeg, with one developer in the loop, Crustify wrapped \lgApiCov\% and \ffApiCov\% of their public APIs; on the interface that consumers are expected to use, only \lgUnsafeFnOtherPct\% and \ffUnsafeFnOtherPct\% of their public functions are \texttt{unsafe}. We plan to extend Crustify to C++ and to incremental ports.

\section*{Data-Availability Statement}
\label{sec:data-availability}

We will release as open source the Crustify harness, including its agent prompts and skills, the \texttt{ffibox}, Wavefront, and ffimap tools, the safe bindings generated for libgit2 and ffmpeg, and the campaign data from which our evaluation is computed, i.e., schedules, ownership stores, review reports, ffimap scans, cost records, and analysis scripts (\url{https://github.com/crustify-rs}). Its main limitation is that rerunning a campaign requires access to the commercial \glspl{llm} we used, costs about \$\lgCostUSD{} for libgit2 and \$\ffCostUSD{} for ffmpeg, and, since \gls{llm} agents are not deterministic, yields bindings that differ from ours; the released campaign data and generated bindings, however, allow every figure in the evaluation to be reproduced without rerunning the agents.

\section*{Use of Generative AI}
The \gls{llm} agents that generate and review the bindings are part of the research method (\S\ref{sec:agents}). In addition, we used \glspl{llm} to assist in editing and revising the text of this paper, checking its claims about prior work against the cited papers, and developing the tools it presents; the authors reviewed all such content and take full responsibility for it.

\begin{acks}
This material is based upon work supported by the Defense Advanced Research Projects
Agency under Agreement HR00112590134.
\end{acks}

\bibliographystyle{ACM-Reference-Format}
\bibliography{references}

\clearpage
\appendix
\emergencystretch=2em
\section{Binding Examples}
\label{sec:appendix-examples}

This appendix illustrates the binding steps of \S\ref{sec:synthesis} with bindings from our campaigns, and maps each to the \texttt{ffibox} primitives of \S\ref{sec:ffibox} that it uses (Table~\ref{tab:examples}).

\begin{table}
  \caption{Examples of C types and functions and the bindings that Crustify emits for them.}
  \label{tab:examples}
  \small
  \begin{tabular}{@{}ll>{\raggedright\arraybackslash}p{0.29\linewidth}@{}}
    \toprule
    C item & Rust binding & \texttt{ffibox} primitive \\
    \midrule
    \texttt{AVBPrint} & \texttt{BPrintBox}, \texttt{BPrintOnBuffer<'b>} & \texttt{CBox} with two policies \\
    \texttt{git\_object} & \texttt{ObjectArc<'r>} & \texttt{CArc}, reference-counted by C \\
    \texttt{av\_file\_map} & \texttt{FileMap} & \texttt{CVec}, length passed to its policy \\
    \midrule
    \texttt{git\_checkout\_options} & \texttt{CheckoutOptions<'o>} & layout with a hand-written lifetime \\
    \texttt{git\_commit} & \texttt{CommitRef<'a, 'r>} & handle with two lifetimes \\
    \midrule
    \texttt{git\_strarray} (result) & \texttt{StrArrayVal} & \texttt{CVal}, disposed on drop \\
    \texttt{git\_strarray} (argument) & \texttt{StrArrayView<'a>} & borrowed view over Rust strings \\
    \texttt{av\_dict\_set} & \texttt{Dict::set}, \texttt{Dict::set\_owned} & forked: borrowed \texttt{\&CStr} or owned \texttt{HeapString} \\
    \midrule
    \texttt{av\_malloc}, \texttt{av\_free} & \texttt{HeapBox<T>}, \texttt{HeapVec<T>} & \texttt{CBox}, \texttt{CVec} sharing one policy \\
    \texttt{git\_odb\_backend\_data\_alloc} & \texttt{OdbDataBox<T>}, \texttt{OdbDataVec<T>} & \texttt{CBox}, \texttt{CVec} with separate policies \\
    \texttt{av\_strdup}, \texttt{av\_free} & \texttt{HeapString} & \texttt{CStrBox} \\
    \texttt{av\_refstruct\_ref} & \texttt{RefStructBox<T>}, \texttt{RefStructArc<T>} & \texttt{CBox}, \texttt{CArc}; \texttt{CZeroable} or \texttt{MaybeUninit<T>} \\
    \midrule
    \texttt{FF\_PAD\_STRUCTURE} & \texttt{PaddedStruct} & sealed trait over layouts \\
    \texttt{AVChannelLayout} & \texttt{LayoutDetails<'a>} & tagged enum with a \texttt{CSlice} variant \\
    \midrule
    \texttt{av\_aes\_size} & \texttt{aes::context\_size()} & \texttt{\&'static} shared reference \\
    \bottomrule
  \end{tabular}
\end{table}

\paragraph{Policies.}
A policy names the C routine that releases an object and, where they exist, those that clone and allocate it, and the owner it parameterizes calls them on drop, on clone, and on construction (C2, \S\ref{sec:ffibox}). Since the routines are bound to the policy rather than to the type, a type can have several. ffmpeg's \texttt{AVBPrint} is owned either as a \texttt{BPrintBox}, whose policy finalizes the buffer's own string with \texttt{av\_bprint\_finalize} and then frees the structure, or as a \texttt{BPrintOnBuffer<'b>}, which prints into a caller's buffer and whose policy frees only the structure, since finalizing it would free the caller's buffer, which the owner keeps borrowed for \texttt{'b}. Both are \texttt{CBox}es. A reference-counted object is held by a \texttt{CArc}, whose policy also clones by reference count (\texttt{CRefClone}, C6): libgit2's \texttt{ObjectArc<'r>} gives back its reference with \texttt{git\_object\_free} on drop and takes another with \texttt{git\_object\_dup} on \texttt{clone}. It never hands out the exclusive handle (C4), since the repository's object cache holds a reference that no owner sees, so the count never proves an owner the only one. An array is held by a \texttt{CVec}, which passes its length to its policy (C5): ffmpeg's \texttt{FileMap}, the contents of a file that \texttt{av\_file\_map} maps, is a \texttt{CVec<u8, FileUnmap>}, whose policy calls \texttt{av\_file\_unmap}, which takes the mapping's size. None of the policies in our campaigns carries run-time state.

\paragraph{Lifetimes.}
Handles carry the lifetime of the memory they reference (C3, \S\ref{sec:ffibox}), and a layout type whose fields borrow memory carries that memory's lifetime too. libgit2's \texttt{git\_checkout\_options} becomes \texttt{CheckoutOptions<'o>}, whose labels, target directory, and baseline tree or index are borrowed from the caller for \texttt{'o}; its label setters, e.g., take an \texttt{Option<\&'o CStr>}. A \texttt{CommitRef<'a, 'r>} borrows a commit for \texttt{'a}, while \texttt{'r} is the borrow of the repository that the commit was looked up in. Its \texttt{message} method returns a \texttt{\&'a CStr}, bound to the commit (\S\ref{sec:synthesis}), whereas lifetime elision would bind the message to the borrow of the handle, so that it could not outlive the handle it was read through. \texttt{ffibox}'s \texttt{define\_ctype!} macro emits only layout types and handles without further lifetime parameters, so these are written by hand, as are all \lgCtypeManual{} hand-written layout types of the libgit2 bindings (\S\ref{sec:eval-guarantees}).

\paragraph{Forked pointers.}
A pointer whose semantics depend on how a function uses it is forked into one Rust form per semantics (\S\ref{sec:synthesis}). The strings of libgit2's \texttt{git\_strarray} are owned by the array when \texttt{git\_tag\_list} or a similar function fills it with a result, which the caller must release with \texttt{git\_strarray\_dispose}, but only borrowed when the caller passes it as an argument, e.g., as a pathspec. A result is therefore held as a \texttt{StrArrayVal}, i.e., a \texttt{CVal} that disposes of the array on drop (C2), whereas an argument is built as a \texttt{StrArrayView<'a>}, which borrows Rust strings for \texttt{'a} and hands out only the shared handle, so that nothing can replace or free them. ffmpeg's \texttt{av\_dict\_set} copies the key and value strings it receives, unless the caller sets \texttt{AV\_DICT\_DONT\_STRDUP\_KEY} or \texttt{AV\_DICT\_DONT\_STRDUP\_VAL}, in which case the dictionary takes ownership of them. It is forked into \texttt{Dict::set}, which borrows its strings as \texttt{\&CStr}, and \texttt{Dict::set\_owned}, which takes them as \texttt{HeapString}s, i.e., strings in ffmpeg's allocator (see below), and sets the flags itself, so that the flags no longer appear in the safe API.

\paragraph{Foundational modules.}
The foundational modules (\S\ref{sec:planning}) bind the routines that allocate, release, and clone type-erased memory and strings as policies, which the other modules' bindings reuse (\S\ref{sec:synthesis}). In ffmpeg, these are \texttt{mem.c}, \texttt{refstruct.c}, and \texttt{avstring.c}. The \texttt{Heap} policy binds \texttt{av\_malloc}, \texttt{av\_mallocz}, \texttt{av\_free}, and \texttt{av\_memdup}, and backs both \texttt{HeapBox<T>}, a \texttt{CBox}, and \texttt{HeapVec<T>}, a \texttt{CVec}, so that single objects and arrays have distinct owners but share one policy. libgit2's public API exports a single raw allocator, \texttt{git\_odb\_backend\_data\_alloc} and \texttt{git\_odb\_backend\_data\_free}, whose bindings instead give single objects and arrays separate policies, behind \texttt{OdbDataBox<T>} and \texttt{OdbDataVec<T>}. Strings, which C allocates by their content's length rather than by a type's size, get policies of their own: ffmpeg's \texttt{HeapString} is a \texttt{CStrBox} that \texttt{av\_free} releases and \texttt{av\_strdup} clones, with safe constructors over \texttt{av\_strdup} and \texttt{av\_strndup}, and the ownership store records functions such as \texttt{av\_asprintf} as string constructors (\S\ref{sec:wavefront}); libgit2's counterpart is \texttt{OdbDataStr}. ffmpeg's refstruct API allocates a reference-counted object of a size and type that the caller chooses, so its policy, which binds \texttt{av\_refstruct\_unref}, \texttt{av\_refstruct\_ref}, and \texttt{av\_refstruct\_exclusive}, is generic over the object's type, and backs both a sole owner, \texttt{RefStructBox<T>}, and a shared one, \texttt{RefStructArc<T>}, which hands out the exclusive handle only when \texttt{av\_refstruct\_exclusive} proves its reference the only one (C4). Its constructors return an initialized object only for types that implement \texttt{CZeroable} (\texttt{av\_refstruct\_allocz}), and \texttt{MaybeUninit<T>} storage otherwise (\texttt{av\_refstruct\_alloc\_ext\_c}).

\paragraph{Generic types and tagged unions.}
Families of related types and type-generating macros are bound as generic abstractions (\S\ref{sec:synthesis}). ffmpeg's \texttt{FF\_PAD\_STRUCTURE} pads a structure to a fixed size, so that later versions can add fields without changing its size. Crustify binds it as a \texttt{PaddedStruct} trait, sealed so that only the macro's instances, such as \texttt{AVBPrint}, implement it, rather than as a generic layout, which would place the padding after the fields' trailing alignment instead of where C puts it; C uses those bytes, e.g., \texttt{AVBPrint}'s internal buffer extends into the padding. A union is bound with the struct that holds its discriminator, as a Rust enum tagged by it (\S\ref{sec:synthesis}): \texttt{AVChannelLayout}'s \texttt{order} field selects between a channel mask and a custom channel map in its union \texttt{u}, which Crustify exposes as a \texttt{LayoutDetails<'a>} enum with one variant per order, e.g., \texttt{Native(u64)} and \texttt{Custom(CSlice<'a, ChannelCustom>)}.

\paragraph{Globals.}
The public APIs of both targets export only immutable globals, which Crustify exposes as \texttt{'static} shared references (\S\ref{sec:synthesis}), e.g., ffmpeg's \texttt{av\_aes\_size}, the size of an AES context, as \texttt{aes::context\_size()}, which returns a \texttt{\&'static i32}.

\section{Cost Analysis}
\label{sec:appendix-batches}

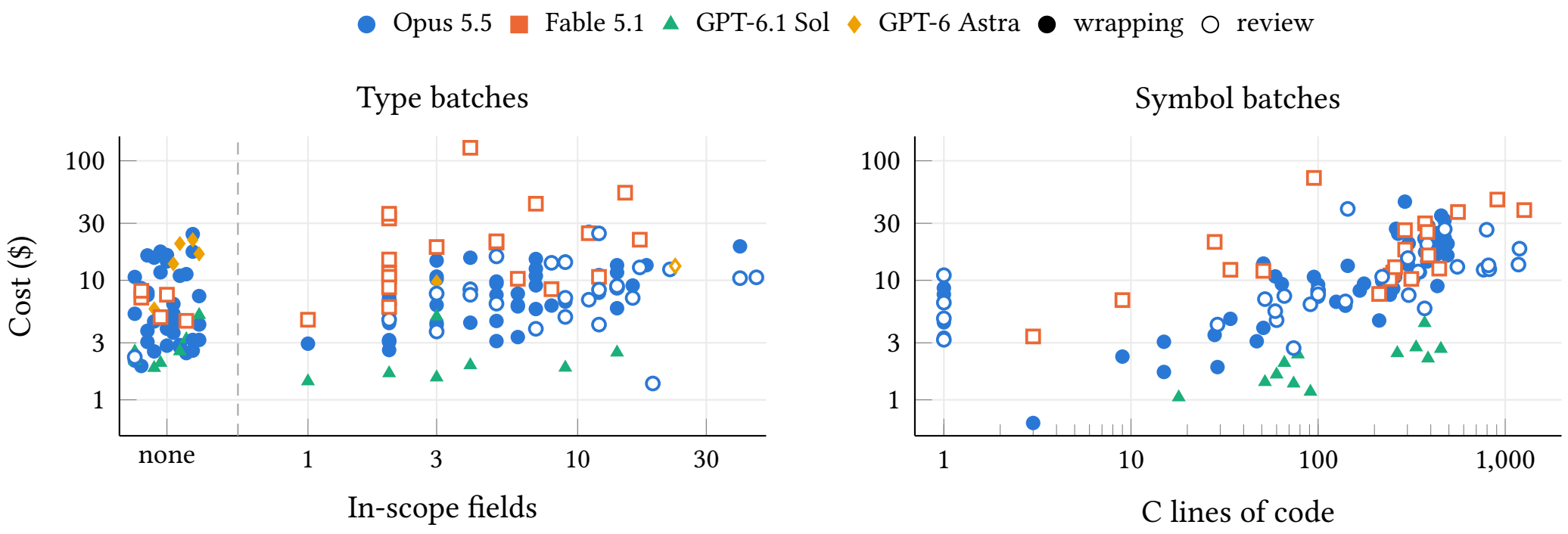
\begin{figure}
  \centering
  \begin{tikzpicture}
  \begin{groupplot}[group style={group size=2 by 1, horizontal sep=1.3cm},
                    batchplot, height=4.2cm,
                    ymode=log, log ticks with fixed point, ytick={1,3,10,30,100},
                    ymin=0.5, ymax=160]
    \nextgroupplot[title={Type batches}, ylabel={Cost (\$)}, xlabel={In-scope fields},
                   xmode=log,
                   xmin=0.2, xmax=50, xtick={0.3,1,3,10,30}, xticklabels={none,1,3,10,30},
                   batch legend, legend to name=batchlegend-size]
      \batchlegend
      \addlegendimage{only marks, mark=*, draw=black, fill=black}\addlegendentry{wrapping}
      \addlegendimage{only marks, mark=*, draw=black, fill=white}\addlegendentry{review}
      \draw[black!35, densely dashed] (axis cs:0.55,0.5) -- (axis cs:0.55,160);
      \batchsize{type}
    \nextgroupplot[title={Symbol batches}, xlabel={C lines of code}, xmode=log,
                   log ticks with fixed point, xtick={1,10,100,1000}, xmin=0.7, xmax=2000]
      \batchsize{symbol}
  \end{groupplot}
  \node[anchor=south] at ($(group c1r1.north)!0.5!(group c2r1.north) + (0,4ex)$) {\pgfplotslegendfromname{batchlegend-size}};
\end{tikzpicture}
  \caption{Cost of each wrapping (filled) and review (hollow) batch of both campaigns against what it wrapped or reviewed: the in-scope fields of its types, with batches of types without in-scope fields, i.e., opaque types and enums, in a separate column, and the C lines of code of its symbols. Both campaigns are pooled; color and marker give the model. The foundational batches, which bind allocation and string routines as lifecycle policies rather than API types and symbols, and one libgit2 review batch, the only one that ran on Opus~4.8, are omitted. All axes are logarithmic.}
  \Description{Two log-log scatter plots of batch cost in dollars, for type batches against their in-scope fields and for symbol batches against their C lines of code. Type-batch cost barely grows with the number of fields, whereas symbol-batch cost grows with the C code wrapped; GPT-6.1 Sol batches lie lowest.}
  \label{fig:cost-size}
\end{figure}

This appendix extends the cost analysis of \S\ref{sec:eval-cost}. Figure~\ref{fig:cost-size} relates the cost of each batch to what it wrapped or reviewed.

\paragraph{Wrapping.}
The cost of a type batch depends little on the number of fields it wraps (Spearman's $\rho = \rhoTypeFields$, and $\rho = \rhoTypeFieldsT$ over the batches that wrap fields), and a batch of types without in-scope fields costs about as much as one with fields (median \$\medTypeNoFields{} versus \$\medTypeFields{}). Wrapping a type thus involves more than its field accessors, likely its lifecycle routines, the uses of the type in the C code that the agent inspects to establish ownership, and its pointer fields, whose representation the agent must decide. A type cost \$\usdPerType{} on average, between \$\usdPerTypeMin{} and \$\usdPerTypeMax{}. The cost of a symbol batch, in contrast, grows with the C code it wraps ($\rho = \rhoSymLoc$): a symbol cost \$\usdPerSym{} on average, or about \$\usdPerKloc{} per thousand lines of C. The model matters as much as the size: the batches that GPT-6.1 Sol wrapped cost \$\usdPerTypeSol{} per type and \$\usdPerKlocSol{} per thousand lines, against \$\usdPerTypeOpus{} and \$\usdPerKlocOpus{} for Opus~5.5. Counting the reviews, which Opus~5.5 performed for the two sub-campaigns that GPT-6.1 Sol wrapped, libgit2's \texttt{config} and ffmpeg's \texttt{crypto}, an item that GPT-6.1 Sol wrapped still cost less, \$\lgSolTotItem{} in libgit2 and \$\ffSolTotItem{} in ffmpeg, against \$\lgOpusTotItem{} and \$\ffOpusTotItem{} for one that Opus~5.5 wrapped. Since the two sub-campaigns differ in their code as well as in their model, these figures do not isolate the model's effect on cost.

\paragraph{Review.}
Review accounts for \lgReviewShare\% and \ffReviewShare\% of the agents' cost, and \fableReviewShare\% of it went to Fable~5.1, which reviewed the early sub-campaigns. Fable cost \$\usdPerItemFable{} per reviewed type or symbol, against \$\usdPerItemOpus{} for Opus~5.5, without finding more: it filed \safePerHundredFable{} \gls{ub} or unsafe reports per \$100 of review, against \safePerHundredOpus{} for Opus~5.5. Since Fable reviewed earlier sub-campaigns, before most of the corrections of the setup, this comparison does not isolate the model. Reviewing whole waves instead of single batches reduced the cost per reviewed item in libgit2 from \$\lgUsdPerItemBatchRv{} to \$\lgUsdPerItemWaveRv{}, and raised the \gls{ub} and unsafe reports per \$100 of review from \lgSafePerHundredBatchRv{} to \lgSafePerHundredWaveRv{}, although only \lgBatchRvCount{} reviews covered single batches; in ffmpeg, whose early reviews already covered several batches, the cost per item did not change.

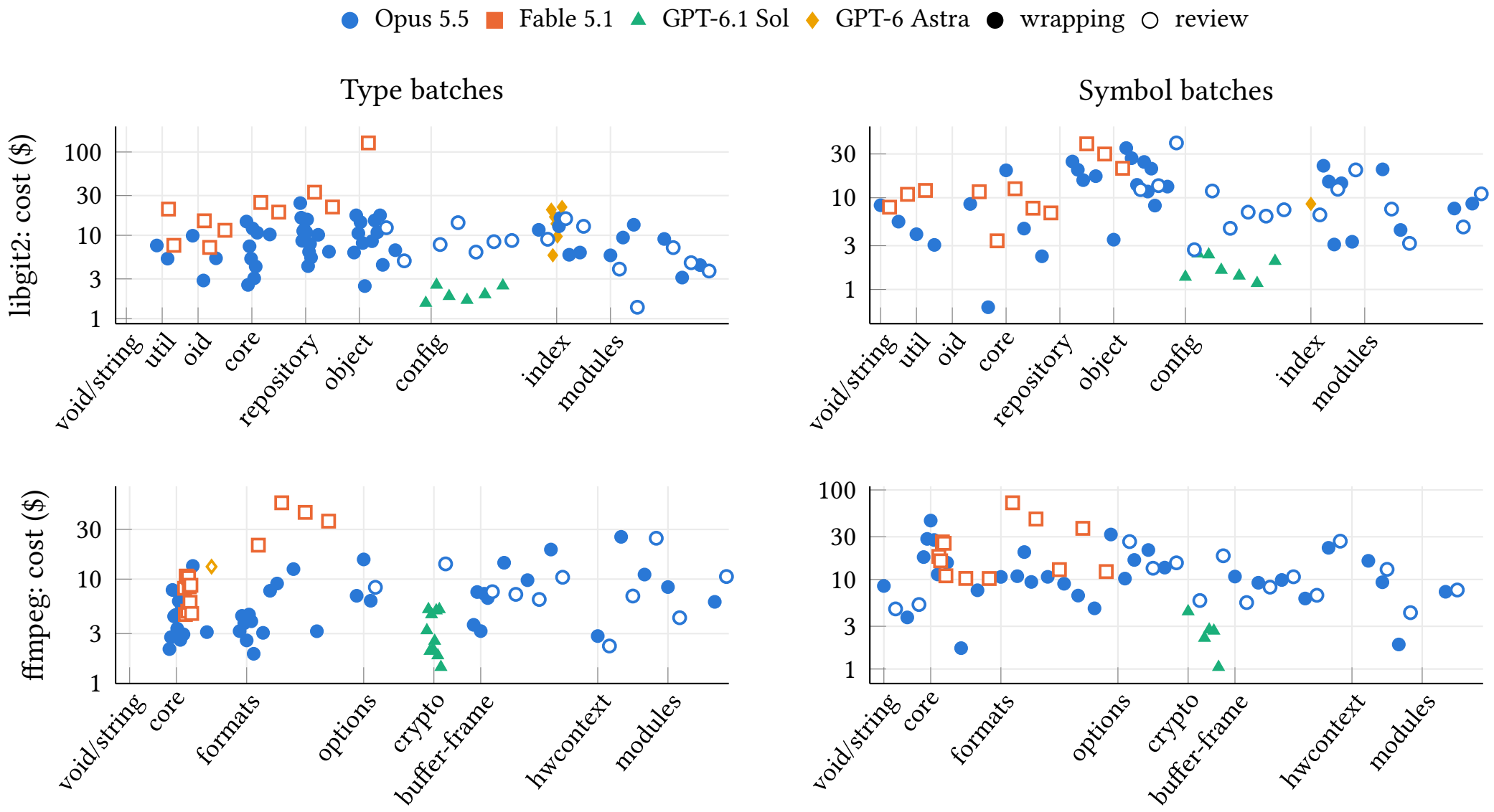
\begin{figure}
  \centering
  \begin{tikzpicture}
  \begin{groupplot}[group style={group size=2 by 2, horizontal sep=1.3cm, vertical sep=1.5cm},
                    batchplot, ymode=log, log ticks with fixed point, ytick={1,3,10,30,100}]
    \nextgroupplot[libgit2 campaign axis, title={Type batches}, ylabel={libgit2: cost (\$)},
                   batch legend, legend to name=batchlegend-cost]
      \batchlegend
      \addlegendimage{only marks, mark=*, draw=black, fill=black}\addlegendentry{wrapping}
      \addlegendimage{only marks, mark=*, draw=black, fill=white}\addlegendentry{review}
      \batchcost{data/libgit2-wrap.csv}{data/libgit2-review.csv}{type}
    \nextgroupplot[libgit2 campaign axis, title={Symbol batches}]
      \batchcost{data/libgit2-wrap.csv}{data/libgit2-review.csv}{symbol}
    \nextgroupplot[ffmpeg campaign axis, ylabel={ffmpeg: cost (\$)}]
      \batchcost{data/ffmpeg-wrap.csv}{data/ffmpeg-review.csv}{type}
    \nextgroupplot[ffmpeg campaign axis]
      \batchcost{data/ffmpeg-wrap.csv}{data/ffmpeg-review.csv}{symbol}
  \end{groupplot}
  \node[anchor=south] at ($(group c1r1.north)!0.5!(group c2r1.north) + (0,4ex)$) {\pgfplotslegendfromname{batchlegend-cost}};
\end{tikzpicture}
  \caption{Cost of each wrapping (filled) and review (hollow) batch in campaign order: the x-axis runs over the campaign's waves, labeled where each sub-campaign starts, the foundational \emph{void} and \emph{string} first, and where the per-module waves that followed them start (\emph{modules}); the batches of one wave, which ran concurrently, are spread around its position, and those of its review wave half a wave later. Color and marker give the model; the libgit2 review batch that ran on Opus~4.8 is omitted. The y-axes are logarithmic.}
  \Description{Four scatter plots of batch cost in dollars on a logarithmic axis against campaign order, for libgit2 and ffmpeg type and symbol batches, with ticks where each sub-campaign starts. Cost follows the sub-campaign and model rather than the position in the campaign.}
  \label{fig:cost}
\end{figure}

\paragraph{Over time.}
Figure~\ref{fig:cost} shows the cost of each batch over the course of both campaigns. The cost of a wrapping batch follows its sub-campaign and model rather than its position in the campaign. In libgit2, the batches of the first sub-campaigns (\texttt{util}, \texttt{oid}, and \texttt{core}) cost a median of \$4.6 to \$6.3, against \$11 to \$13 for the larger \texttt{repository}, \texttt{object}, and \texttt{index}, and the two sub-campaigns that GPT-6.1 Sol wrapped, libgit2's \texttt{config} and ffmpeg's \texttt{crypto}, are the cheapest of their campaigns, with medians of \$1.9 and \$2.7 (\S\ref{sec:eval-setup}). The highest median costs of review batches, \$31 in libgit2's \texttt{repository} and \$37 in ffmpeg's \texttt{formats}, both come from sub-campaigns that Fable~5.1 reviewed.

\section{Comparison of ffimap with Unsafe-Code Scanners}
\label{sec:appendix-ffimap}

We compare ffimap (\S\ref{sec:scan}) with three tools that measure \texttt{unsafe} code in Rust crates: cargo-geiger~\cite{cargogeiger} (0.14.0), which parses every source file of a package and counts its \texttt{unsafe} items and expressions; cargo-scan~\cite{zoghbi2026cargoscan}, which parses the source with name resolution from rust-analyzer and reports effects, such as \gls{ffi} calls and raw-pointer dereferences, per call site, optionally after expanding macros (\texttt{-e}, a flag of its implementation~\cite{cargoscantool}); and Clippy~\cite{clippy} (0.1.99), restricted to its lints on \texttt{unsafe} code and pointer casts. All four scan the two crates over libgit2 that Table~\ref{tab:baselines} compares: our libgit2 bindings at the campaign's tip (\scOurLines{} lines of code) and git2-rs~0.21.0 (\scGitLines{} lines). We measure only the library crate, excluding \texttt{-sys} crates and tests. Table~\ref{tab:scanners} summarizes the counts.

\begin{table}
  \caption{Counts of ffimap and other scanners on our libgit2 bindings and git2-rs. \emph{scan} denotes cargo-scan, and \emph{scan -e} cargo-scan with macro expansion.}
  \label{tab:scanners}
  \small
  \begin{tabular}{@{}llrr@{}}
    \toprule
    Metric & Tool & Crustify & git2-rs \\
    \midrule
    \texttt{unsafe fn} & ffimap / geiger & \scOurFfimapUnsafeFn{} / \scOurGeigerUnsafeFn & \scGitFfimapUnsafeFn{} / \scGitGeigerUnsafeFn \\
    \texttt{unsafe impl} & ffimap / geiger / scan & \scOurFfimapUnsafeImpl{} / \scOurGeigerUnsafeImpl{} / \scOurScanUnsafeImpl & \scGitFfimapUnsafeImpl{} / \scGitGeigerUnsafeImpl{} / \scGitScanUnsafeImpl \\
    \midrule
    \texttt{unsafe} expressions & geiger & \scOurGeigerUnsafeExpr & \scGitGeigerUnsafeExpr \\
    \texttt{unsafe} blocks & ffimap & \scOurFfimapUnsafeBlock & \scGitFfimapUnsafeBlock \\
    Lines in \texttt{unsafe} blocks & ffimap & \scOurFfimapUnsafeLines & \scGitFfimapUnsafeLines \\
    Blocks with several \texttt{unsafe} operations & Clippy & \scOurClippyMultiOps & \scGitClippyMultiOps \\
    \midrule
    FFI calls & ffimap / scan / scan -e & \scOurFfimapFfiCall{} / \scOurScanFfiCall{} / \scOurScanEFfiCall & \scGitFfimapFfiCall{} / \scGitScanFfiCall{} / \scGitScanEFfiCall \\
    Raw-pointer dereferences & ffimap / scan & \scOurFfimapPtrDeref{} / \scOurScanPtrDeref & \scGitFfimapPtrDeref{} / \scGitScanPtrDeref \\
    \bottomrule
  \end{tabular}
\end{table}

\paragraph{Unsafe items.}
On git2-rs, ffimap and cargo-geiger agree to within one item. On our bindings, cargo-geiger's \scOurGeigerUnsafeImpl{} \texttt{unsafe} implementations are exactly those that the source spells out: it misses the \scGeigerMissImpl{} (\scGeigerMissImplPct\%) that \texttt{ffibox}'s macros generate, e.g., \texttt{define\_ctype!}, since a scanner that parses source text does not see items that macros generate. Its \scOurGeigerUnsafeFn{} \texttt{unsafe} functions likewise leave out the \scOurMacroUnsafeFn{} that \texttt{ffibox}'s policy macros generate, but include items that ffimap does not count as function definitions, such as \texttt{unsafe} trait-method declarations. cargo-scan reports the same \scOurScanUnsafeImpl{} \texttt{unsafe} implementations as cargo-geiger, with or without \texttt{-e}.

\paragraph{Unsafe volume.}
Our bindings follow a convention of one narrow \texttt{unsafe} block per operation, which their lints enforce. They therefore contain \scBlocksRatio{} times as many \texttt{unsafe} blocks as git2-rs, so that cargo-geiger's expression count rates them as \scGeigerExprMorePct\% more \texttt{unsafe}, but their blocks span \scLinesFewerPct\% fewer lines, and Clippy reports no block with several \texttt{unsafe} operations, against \scGitClippyMultiOps{} in git2-rs.

\paragraph{FFI calls.}
git2-rs routes most of its \gls{ffi} calls through a macro, so cargo-scan finds \scGitScanFfiCall{} of them without expansion and \scGitScanEFfiCall{} with it, within \scGitScanEFfiDiffPct\% of ffimap's \scGitFfimapFfiCall. On our bindings, whose \texttt{-sys} crate keeps its generated declarations in its source tree, cargo-scan finds \scOurScanFfiCall{} and \scOurScanEFfiCall{} of ffimap's \scOurFfimapFfiCall{} (\scOurScanFfiPct\% and \scOurScanEFfiPct\%), but only \scOurScanPtrDeref{} of its \scOurFfimapPtrDeref{} raw-pointer dereferences. The dereferences likely sit mostly in the raw borrows through which the bindings project fields, e.g., \texttt{\&raw const (*p).field}: cargo-scan's parser rejects this Rust 2024 syntax and skips each such expression with a warning.

\paragraph{API surface.}
Only ffimap separates the public interface from the implementation; the other tools count per package, file, or function. Table~\ref{tab:baselines} compares the two crates' public interfaces as ffimap measures them.

\paragraph{Run time.}
Table~\ref{tab:scanner-time} gives the wall time of each tool, cold with an empty target directory and warm as the median of three subsequent runs, on the machine of \S\ref{sec:eval-setup}, which another workload shared, with a load average between \scLoadMin{} and \scLoadMax{} of its 144 hardware threads. ffimap forces a recompilation on each run, since a cached build emits no measurements, so its warm runs re-measure the crate. On git2-rs, it does so \scGitFfimapVsGeiger{} times faster than cargo-geiger, which re-parses every source file on each run, but slightly slower on our bindings (\scOurFfimapWarm{} against \scOurGeigerWarm{}~s), at \scOurFfimapRssShare{} and \scGitFfimapRssShare{} of cargo-geiger's memory. Clippy's warm runs replay cached diagnostics and therefore analyze little. cargo-scan's cold runs are the slowest, presumably because rust-analyzer loads the workspace and runs its build scripts, which for git2-rs compile the vendored libgit2 sources.

\begin{table}
  \caption{Wall time of each scanner, cold and warm, and peak memory of its warm runs.}
  \label{tab:scanner-time}
  \small
  \begin{tabular}{@{}lrrrr@{}}
    \toprule
    & \multicolumn{2}{c}{Cold / warm (s)} & \multicolumn{2}{c}{Peak memory (MB)} \\
    Tool & Crustify & git2-rs & Crustify & git2-rs \\
    \midrule
    ffimap & \scOurFfimapCold{} / \scOurFfimapWarm & \scGitFfimapCold{} / \scGitFfimapWarm & \scOurFfimapRss & \scGitFfimapRss \\
    cargo-geiger & \scOurGeigerCold{} / \scOurGeigerWarm & \scGitGeigerCold{} / \scGitGeigerWarm & \scOurGeigerRss & \scGitGeigerRss \\
    cargo-scan & \scOurScanCold{} / \scOurScanWarm & \scGitScanCold{} / \scGitScanWarm & \scOurScanRss & \scGitScanRss \\
    cargo-scan \texttt{-e} & \scOurScanECold{} / \scOurScanEWarm & \scGitScanECold{} / \scGitScanEWarm & \scOurScanERss & \scGitScanERss \\
    Clippy & \scOurClippyCold{} / \scOurClippyWarm & \scGitClippyCold{} / \scGitClippyWarm & \scOurClippyRss & \scGitClippyRss \\
    \bottomrule
  \end{tabular}
\end{table}

\paragraph{Limitations.}
cargo-scan requires a literal package version, and otherwise reports no effects while exiting successfully; we therefore ran all tools on a copy of our bindings whose manifest states its version, on which ffimap reports the same counts as on the original. We did not run Siderophile or Rudra, which require older pinned toolchains and do not count the \texttt{unsafe} surface. Each crate was measured at a single commit.

\section{Wavefront Evaluation Details}
\label{sec:appendix-wavefront}

\paragraph{Setup.}
We measured Wavefront at commit \texttt{6150082} on the relations extracted from both campaigns' C revisions (libgit2 \texttt{0551dfd}, ffmpeg \texttt{1019f8f}), with the campaign-wide configurations from which their current plans were made, whose closure holds the files of the campaigns' modules. We report two graph modes: the \emph{implementation} graph, which follows function bodies and corresponds to what translation pipelines schedule, and the \emph{public-signature} graph, from which both campaigns are scheduled (\S\ref{sec:wavefront}, D4). Table~\ref{tab:wf-graphs} gives their properties.

\paragraph{Cost.}
Wavefront's setup runs once per target. With CodeQL~2.26.3, each campaign's build recipe, and a machine with 144 hardware threads shared with other workloads (load average between 16 and 47), configuring each target took 7\,s for libgit2 and 18\,s for ffmpeg, building the CodeQL database from a traced build 120\,s and 328\,s, and extracting its relations with Wavefront's 25 queries 847\,s and 1{,}034\,s, i.e., 34\,s and 41\,s per query, since each query runs as a separate CodeQL process with a fixed startup cost. Scheduling each campaign as a whole from the extracted relations, with a cold cache and thus including the composition of its public-signature graph, then takes 33\,s and 54\,s, for about 17 and 24 minutes in all. During the campaigns, agents' queries took a median of 5.0\,s on libgit2 and 37.9\,s on ffmpeg, since Wavefront recomposes its records from the extracted relations on every call, whereas cached graph queries took a median of 0.28\,s and 1.5\,s.

\begin{table}
  \caption{Dependency graphs of both campaigns' closures in implementation (impl) and public-signature (sig) mode.}
  \label{tab:wf-graphs}
  \small
  \begin{tabular}{@{}lrrrr@{}}
    \toprule
    & libgit2 impl & libgit2 sig & ffmpeg impl & ffmpeg sig \\
    \midrule
    Nodes & 7{,}545 & 7{,}526 & 5{,}777 & 5{,}651 \\
    Edges & 49{,}528 & 7{,}436 & 21{,}259 & 3{,}400 \\
    Non-trivial SCCs & 22 & 7 & 19 & 4 \\
    Largest SCC (nodes) & 135 & 8 & 21 & 5 \\
    Nodes in cycles & 3.35\% & 0.39\% & 1.47\% & 0.19\% \\
    Cut edges & 126 & 22 & 43 & 6 \\
    Layers (longest chain after cuts) & 53 & 7 & 33 & 10 \\
    Depth with collapsed SCCs & 66 & 8 & 32 & 10 \\
    \bottomrule
  \end{tabular}
\end{table}

\paragraph{Determinism.}
Regenerating each campaign's schedule under three hash seeds, with a cold and a warm cache, gives byte-identical schedules, and regenerating it from extracted relations whose rows are shuffled places every item in the same wave and batch, with the same record. libgit2's schedule is byte-identical to the one that its planner generated; ffmpeg's differs by one unit, the system's \texttt{struct tm}, which the planner's earlier version of Wavefront scheduled before a fix kept forward-declared external tags out of the units. An earlier evaluation exposed an implementation bug that these runs no longer show: Wavefront resolved struct tags that several C files define privately by the order of its extracted relations, which added wrong edges to the graphs from which the campaigns were first scheduled, 214 in ffmpeg and one in libgit2.

\paragraph{Reproducibility.}
Rebuilding both CodeQL databases with each campaign's build recipe and extracting the relations anew, as timed above, reproduces both campaigns' schedules: every item is placed in the same wave and batch.

\section{Comparison with Ownership Inference}
\label{sec:appendix-ownership}

\paragraph{\&inator.}
\&inator~\cite{chen2026inator} is conceptually closest to Crustify's ownership analysis: it infers an ownership and borrowing discipline for the interface of a C program, i.e., its struct fields and function signatures, and expresses it as Rust types, by encoding semantic equivalence and Rust's borrow-checking rules as constraints that Z3 solves over the whole program. We did not run it on our targets, for three reasons. First, its stated limitations exclude features on which both libraries rely: it provides partial or no support for function pointers, polymorphic pointers, unions, and variadic functions, and does not support multi-threaded programs or libraries; libgit2, e.g., allocates and releases memory through the function pointers of its pluggable allocator. Second, it requires the whole program and, for a library, a client test suite that exercises it, and its constraint solving takes several hours for programs of a few thousand lines, whereas the code bases of libgit2 and ffmpeg comprise about 285 and 1{,}420 thousand lines. Third, its output does not correspond to ours: it selects a single type per pointer, e.g., \texttt{Box<T>} or \texttt{Rc<RefCell<T>>}, for a rewrite into Rust-owned memory, whereas Crustify's bindings keep objects in C memory, release them through the library's own routines, and fork pointers whose ownership depends on the path.

\paragraph{Crown and PtrTrans.}
We compared the ownership records that the agents submitted to Wavefront during the libgit2 campaign, covering 1{,}124 pointer positions in fields, parameters, and returns of the public API, i.e., the \lgOwnPos{} positions of RQ5 restricted to functions that libgit2 exports and to public typedefs, with the static ownership inference of Crown~\cite{zhang2023crown} and of PtrTrans~\cite{ptrtrans2026}, whose SVF-based analysis computes the ownership labels that its \gls{llm} receives. Neither produces usable ownership for libgit2. Crown analyzes C2Rust's translation of libgit2 in 11 minutes, but only after seven adaptations of the input that work around C2Rust and Crown defects and four patches that keep Crown from aborting on unions and on per-function assertion failures, none of which adds information to the analysis. Its constraints are unsatisfiable for 31\% of the 5{,}638 function bodies, it labels only 38\% of the public parameters, 31\% of the returns, and 60\% of the fields, and it does not model function pointers. Where it does label a position, it agrees with the agents on 92\% of the parameters, but only because most API pointers are borrowed and Crown labels nearly every pointer as borrowed: it never labels a public parameter or return as owned, labels all 28 labeled arguments of destructors such as \texttt{git\_repository\_free} as borrowed, and agrees on the ownership of the objects that constructors write into out-parameters for 3 of 51 positions. 

Over the whole program, it labels 48 of about 8{,}100 pointers as owned. PtrTrans's analysis did not finish: after three and a half hours, it had processed 448 of 5{,}741 functions, all leaves of the call graph, and none of the 609 structs, since it scans the whole program's control-flow graph for every pointer parameter. Both tools share a blind spot: they recognize ownership only at calls to the C library's allocation and release functions, e.g., \texttt{malloc}, \texttt{realloc}, and \texttt{free}, which they match by name, whereas libgit2 allocates and releases all memory through the function-pointer table of its pluggable allocator, so that neither sees an allocation or a release in the public API. Crustify's agents instead read this convention from the code, as a developer writing bindings would. Since agreement with the agents' records does not establish correctness, these figures compare the tools rather than measure the agents' accuracy, although the destructor and constructor positions are unambiguous in the C source; we ran the comparison on libgit2, whose allocator neither tool sees, and leave ffmpeg, whose allocator wraps \texttt{malloc}, to future work.

\end{document}